\documentclass[aps,prb,preprint,showpacs,showkeys,nofootinbib]{revtex4-1}
\usepackage[T2A,OT1]{fontenc}
\usepackage[english]{babel}
\usepackage{graphicx} 
\usepackage{latexsym}
\usepackage{amsbsy}         
\usepackage{amsfonts}
\usepackage{eurosym}
\usepackage{xcolor}[2007/01/21]

\begin{document}
\title{Berry-phase effects and electronic dynamics in noncollinear
antiferromagnetic texture}
\author{Olena Gomonay}
\affiliation {
 National Technical University of Ukraine ``KPI''\\ ave Peremogy, 37, 03056, Kyiv,
Ukraine}

\begin{abstract}
Antiferromagnets (AFMs), in contrast to ferromagnets, show a
nontrivial magnetic structure with zero net magnetization. However,
they share a number of spintronic effects with ferromagnets,
including spin-pumping and spin transfer torques. Both phenomena
stem from the coupled dynamics of free carriers and localized
magnetic moments. In the present paper I study the adiabatic dynamics
of a spin-polarized electrons in a metallic AFM exhibiting a noncollinear
120$^\circ$ magnetic structure. I show that the slowly varying AFM
spin texture  produces a non-Abelian gauge potential related to the
time/space gradients of the N\'{e}el vectors. Corresponding
emergent electric and magnetic fields induce rotation of spin and
influence the orbital dynamics of free electrons. I discuss both the possibility
of a topological spin Hall effect in the vicinity of topological AFM
solitons with nonzero curvature and rotation of the electron spin traveling through
the AFM domain wall.
\end{abstract}
\maketitle
\section{Introduction}
Metallic and semiconducting antiferromagnets (AFM) with high
ordering (N\'{e}el) temperature are promising candidates for
spintronic applications. Compared to their ferromagnetic
counterparts, AFM-based devices show reduced critical currents for
magnetization switching \cite{Tang:2009} and can effectively
operate at higher frequencies.\cite{Kampfrath:2011} According to
theoretical predictions, AFMs can also show the following current-induced
phenomena typical for ferromagnets: spin-transfer torques,
\cite{Haney:2008, Gomonay:2008_JMSJ, gomo:2008E} spin pumping
\cite{Cheng:PhysRevLett.113.057601}, domain wall motion
\cite{Tserkovnyak:PhysRevLett.106.107206,
Brataas:PhysRevLett.110.127208}, -- but with much richer physics
stemming from the nontrivial magnetic structure.

However, the mechanisms responsible for the coupled dynamics of free electrons and localized magnetic moments in
AFMs are still not clear and thus recently became a matter of interest.
For example, ferromagnets can work as spin polarizers due to exchange coupling between the localized spins (that contribute to macroscopic  magnetization) and the spin of the conduction electron. In contrast, the AFMs have zero or vanishingly small magnetization. The symmetry properties of the N\'eel vector (AFM order parameter) differ from those of the spin vector and thus the polarization mechanism through $sd$-exchange is excluded. On the other hand, AFMs have a nontrivial magnetic structure which removes degeneracy of otherwise equivalent directions and/or planes and thus can affect spin dynamics of free electron.
The nontrivial spintronic effects in AFMs are usually attributed the
to $sd$-exchange and/or spin-orbit coupling. In homogeneous
systems the spin-orbit interaction can induce polarization of the
electric current which flows through the collinear AFM
\cite{Zelezny:PhysRevLett.113.157201} or the anomalous Hall effect in
noncollinear planar AFM.\cite{MacDonald:PhysRevLett.112.017205,
Kubler:2014arXiv1410.5985K} Exchange interaction itself can induce the
topological Hall effect in the structure with the nonzero chirality
\cite{Bruno:PhysRevLett.93.096806,MacDonald:RevModPhys.82.1539},
e.g. in non-coplanar AFMs.\cite{Takatsu:PhysRevLett.105.137201,Christoph_Surgers:2014}

The $sd$-exchange also plays an important role in the magnetic
textures which can produce Abelian
\cite{Tserkovnyak:PhysRevB.79.014402} and non-Abelian
\cite{Cheng:PhysRevB.86.245118} gauge potentials for conduction
electrons. Corresponding fields contribute to adiabatic
spin-transfer torque and spin-pumping phenomena and thus could be
experimentally detected. In particular, a \emph{ferro}magnetic
texture produces an effective spin-dependent $U$(1) gauge field
for conduction electrons. This gauge field gives rise to 
effective electrical and magnetic fields proportional to the
macroscopic magnetization of the ferromagnet. An alternative point of
view relates the emergent fields with the Berry phase accumulated
by a free electron whose spin is aligned with the local macroscopic
magnetization.\cite{Barnes:PhysRevLett.98.246601} An analogous
influence of the collinear \emph{antiferro}magnetic texture on the
dynamics of free carriers was recently predicted in Ref.~\onlinecite{Cheng:PhysRevB.86.245118} from semiclassical analysis of
Berry phase. The Berry phase also strongly affects the semiclassical
motion of electrons in chiral magnets with spin-orbit interaction
and can even induce formation of skyrmions in these materials
("driving force for formation"), as was recently demonstrated in
Ref.~\onlinecite{Mokrousov:PhysRevB.88.214409}.

The present paper focuses on the dynamics of spin-polarized
electrons in the gauge potentials produced by the space/time varying magnetic
moments of a noncollinear
AFM. The main idea is to demonstrate
that the adiabatic dynamics of free electrons is intimately related to the
``dynamic'' magnetization of AFM (no matter
how complicated AFM structure is) and in this sense is similar to adiabatic
dynamics of transport electrons in ferromagnets. Correspondingly, the curvature of a smooth distribution of AFM
vectors generates effective electric and magnetic fields that affect, via $sd$-exchange coupling, the orbital motion of free electrons. I
predict a possibility of topological spin Hall effect in noncollinear AFM
textures that, in analogy with its ferromagnetic counterpart,
originates from the $sd$-exchange. This effect can be used to generate
spin-currents and probe the curvature of the AFM distribution. I
demonstrate that an inhomogeneous AFM structure induces a rotation of
the spin polarization (analog of Faraday effect). This effect can be an
efficient tool for probing theAFM domain structure by electrical
methods.



\section{Model and formalism}
\begin{figure}[tbp]
\centering
\includegraphics[width=0.5\linewidth]{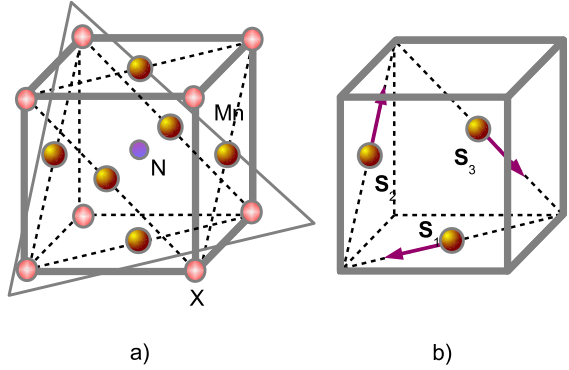} 
\caption{(Color online) Unit cell (a) and $\Gamma^{5g}$ magnetic
structure (b) of antiperovskite Mn$_{3}$XN. Magnetic moments
$\mathbf{S}_1$, $\mathbf{S}_2$ and $\mathbf{S}_3$ are localized at
Mn atoms. } \label{fig_cube_1}
\end{figure}

As a prototype of a conductive antiferromagnet with noncollinear
magnetic structure I consider the antiperovskite Mn$_{3}$XN (X=Ag, Zn,
Ni, Ga) with cubic space group Pm3m (see Fig.\ref{fig_cube_1}) and
with the magnetic moments localized at Mn atoms.\cite{Bertaut1968251}
Alhough some authors \cite{Jardin:1975, Ali2014141,
Motizuki:0022-3719-21-30-011} underline that strong hybridization
around the Fermi level points to the itinerant nature of
antiferromangetism in Mn-based antiperovskites, these substances
could be effectively described by the N\'eel model of
magnetic
sublattices.\cite{Gomonaj:Phase_Tr_1992a,Gomonaj:Phase_Tr_1992,
Lukashev:PhysRevB.82.094417}

The Mn-based perovskites combine the nontrivial triangular magnetic structure 
with the peculiar transport properties. The
linear temperature dependence of the resistivity and strong hybridization of Mn-3$d$ and N-2$p$ electrons around the Fermi level \cite{Ali2014141} suggests a metallic character of conductivity. However, conductivity  and 
temperature coefficient of resistivity are much
smaller than those of a typical metal.\cite{Chi2001307,Takenaka:apl/98/2/10.1063/1.3541449} Hence, these compounds could
be considered as bad metals with hopping character of conductivity.
Resistivity measurements \cite{Takenaka:apl/98/2/10.1063/1.3541449} also
indicate to a strong coupling between the magnetic structure and the transport
properties. In addition, suppressed $sd$ scattering in the magnetic phase
\cite{Takenaka:apl/98/2/10.1063/1.3541449} enables observation of the quantum
phase effects. Thus, Mn$_{3}$XN compounds are generic materials for analysis of the Berry-phase effects in noncollinear AFMs.

\subsection{Magnetic structure formed by localized moments}
The localized magnetic moments (sublattice magnetizations) represented
by three classical vectors $\mathbf{S}_{j}$, $j=1,2,3$ form a
non-collinear
coplanar structure classified\footnote{%
Stricktly speaking, the $\Gamma^{4g}$ structure realized in a certain
temperature range is consistent with weak ferromagnetic structure
in which the vectors $\mathbf{S}_{j}$ slightly deviate from the plane.
We, however, neglect small unncompensated magnetization for the
sake of simplicity.}, depending on the material, as $\Gamma^{5g}$
or $\Gamma ^{4g}$.\cite{Bertaut1968251} Vectors $\mathbf{S}_{j}$
make $120^\circ$ angle with respect to each other. Thus, the
total magnetization cancels within the plane. The ordering plane is defined
by the plane normal $\mathbf{n}$. Well below the N\'eel
temperature $\left\vert \mathbf{S}_{j}\right\vert =S$.

\begin{figure}[tbp]
\centering
\includegraphics[width=0.5\linewidth]{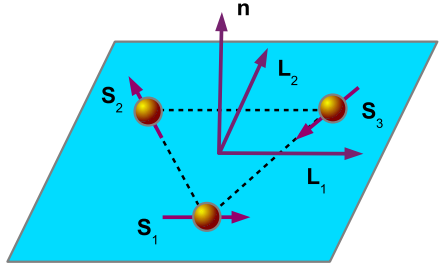}
\caption{(Color online) Local frame generated by AFM vectors
$\mathbf{L}_1$, $\mathbf{L}_2$, and vector $\mathbf{n}$ normal to
the plane of spin ordering.} \label{fig_cube_3}
\end{figure}
Within the macroscopic approach a noncollinear AFM structure can be conveniently described
by two mutually orthogonal AFM, or N\'eel, vectors $%
\mathbf{L}_{1}\perp \mathbf{L}_2$ ($\left\vert \mathbf{L}_{1}\right\vert
=\left\vert \mathbf{L}_{2}\right\vert =S$) that could be considered as a multicomponent order
parameter:
\begin{equation}
\mathbf{L}_{1}=\frac{1}{\sqrt{6}}\left( 2\mathbf{S}_{1}-\mathbf{S}_{2}-%
\mathbf{S}_{3}\right) ,\mathbf{L}_{2}=\frac{1}{\sqrt{2}}\left( \mathbf{S}%
_{2}-\mathbf{S}_{3}\right), \label{eq:AFM_structure_Neel_vectors}
\end{equation}
and the macroscopic magnetization vector
\begin{equation}
\mathbf{M}=\frac{1}{\sqrt{3}}\left( \mathbf{S}%
_{1}+\mathbf{S}_{2}+\mathbf{S}_{3}\right).  \label{eq:AFM_structure_mgnetization}
\end{equation}
From the symmetry point of view vectors
(\ref{eq:AFM_structure_Neel_vectors}) and
(\ref{eq:AFM_structure_mgnetization}) belong to different
irreducible representations of the permutation group $P_3$
(corresponding to the exchange symmetry of the
crystal).\cite{Gomonaj:Phase_Tr_1992a}

In the AFM ground state the macroscopic magnetization $\mathbf{M}%
=0$. Three orthogonal vectors $\mathbf{L}_1\perp \mathbf{L}_2\perp \mathbf{n%
}$ generate a natural local frame for free spin (see Fig.\ref{fig_cube_3}).

In equilibrium homogeneous state the corresponding vectors are $\mathbf{n}^{(0)}\parallel [ 111]$, $%
\mathbf{L}_{1}^{(0)}\Vert[ 0\bar{1}1]$, and  $\mathbf{L}_{2}^{(0)}\Vert
[2\bar{1}\bar{1}]$. In the texture (inhomogeneous state)the orientation
of the sublattice magnetizations can smoothly vary at the length-scale much
greater than interatomic distances, thus, all the magnetic vectors $\mathbf{S}%
_{j}$, $\mathbf{L}_{1}$, $\mathbf{L}_{2}(t,\mathbf{r})$ are continuous
functions of time and space.

Strong exchange coupling locks the mutual orientation of localized moments even
in the presence of relatively small external fields. However, under the
action of these fields the whole structure can smoothly rotate with respect
to some initial configuration (labeled further with the subscript $(0)$). In
the adiabatic approximation which we consider below, the large scale variation of
localized moments is equivalent to solid-like rotation of vectors $\mathbf{S}%
_{j}$ (and, correspondingly, $\mathbf{L}_{1}, \mathbf{L}_{2}, \mathbf{n}$)
and can be conveniently parametrized \cite{Andreev:1980,
Gomonay:PhysRevB.85.134446} with the Gibbs' vector $\boldsymbol{\varphi }$
(so-called Cayley-Gibbs-Rodrigues parametrization) as follows:
\begin{equation}
\mathbf{S}_{j}=\Re (\boldsymbol{\varphi )S}_{j}^{(0)}\equiv \frac{1-%
\boldsymbol{\varphi }^{2}}{1+\boldsymbol{\varphi }^{2}}\mathbf{S}_{j}^{(0)}+%
\frac{2}{1+\boldsymbol{\varphi }^{2}}\left[\boldsymbol{\varphi \times
\mathbf{S}_{j}^{(0)}+\varphi }\left( \boldsymbol{\varphi \cdot \mathbf{S}%
_{j}^{(0)}}\right) \right].  \label{eq:Gibbs_vector}
\end{equation}

Here $\Re (\boldsymbol{\varphi })$ is the orthogonal rotation tensor, $%
\boldsymbol{\varphi }(t,\mathbf{r})\equiv \tan \left( \theta /2\right)
\mathbf{N}$, where $\mathbf{N}(t,\mathbf{r})$ is the instant rotation axis, $%
\theta (t,\mathbf{r})$ is the rotation angle and we treat all the parameters
as continuous functions of space and time.

\begin{figure}[tbp]
\centering
\includegraphics[width=0.5\linewidth]{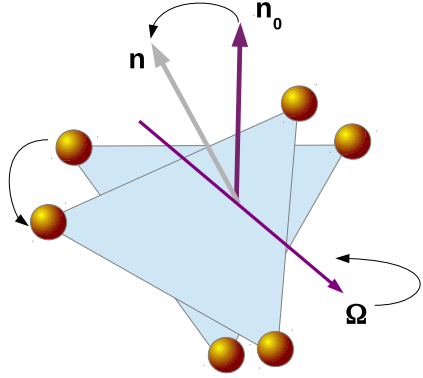}
\caption{(Color online) Solid-like rotation of the magnetic structure. Angular velocity $%
\boldsymbol{\Omega}$ may result either from the time or space rotation of the
AFM structure, as described in the text.}
\label{fig_cube_4}
\end{figure}

Rotation of AFM moments plays an important role in the magnetic dynamics of
localized spin. As it was pointed out by Andreev and Marchenko, \cite%
{Andreev:1980} a solid-like rotation of spins induces nonzero, ``dynamic''
magnetization of AFM, $\mathbf{M}_{dyn}\propto \hat{\chi}\boldsymbol{%
\Omega }_{t}$ ($\hat{\chi}$ is a tensor of magnetic susceptibility), which
is proportional to the pseudovector  $\boldsymbol{\Omega
}_{t}$ of angular velocity, frequently referred to as macroscopic spin (see Fig.\ref{fig_cube_4}%
):
\begin{equation}
\boldsymbol{\Omega }_{t}=2\frac{\partial _{t}\boldsymbol{\varphi} +%
\boldsymbol{\varphi} \times \partial_{t}\boldsymbol{\varphi }}{1+\boldsymbol{%
\varphi }^{2}},~\partial _{t}\Re (\boldsymbol{\varphi )=\boldsymbol{\Omega }}%
_{t}\mathbf{\times }\Re (\boldsymbol{\varphi )}.  \label{eq:time_frequency}
\end{equation}

However, a free electron moving with velocity $\dot{r}_{l}$ ($l=x,y,z$) in
the slowly varying AFM texture should also ``feel'' the effective
magnetization produced by space rotations of the AFM moments and described by the
``space'' angular velocity:
\begin{equation}
\boldsymbol{\Omega }_{l}=2\frac{\partial _{l}\boldsymbol{\varphi +\varphi
\times}\partial_{l}\boldsymbol{\varphi}}{1+\boldsymbol{\varphi }^{2}}%
,~\partial _{l}\Re (\boldsymbol{\varphi )=\boldsymbol{\Omega }}_{l}\mathbf{%
\times }\Re (\boldsymbol{\varphi }).  \label{eq:space_frequency}
\end{equation}

Thus, in the continuous medium the dynamic magnetization of AFM seen by
conduction electron is proportional to the angular velocity $\mathbf{\Omega }%
=\boldsymbol{\Omega }_{t}+\mathbf{\boldsymbol{\Omega }}_{l}\dot{r}_{l}$.

In what follows we consider the functions $\boldsymbol{\varphi}%
(t,\mathbf{r}), \boldsymbol{\Omega}(t,\mathbf{r})$ that describe the AFM texture as given, putting aside
the problem of current-induced dynamics of localized spins.

\subsection{Effective Hamiltonian and band structure}

\begin{figure}[tbp]
\centering
\includegraphics[width=0.5\linewidth]{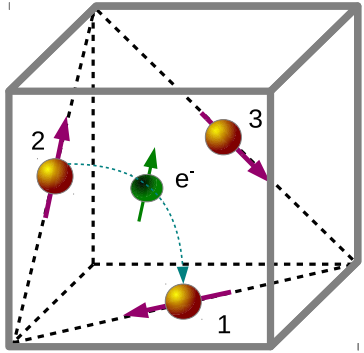}
\caption{(Color online) Electron ($e^-$) hopping between the
magnetic sites. Different sites generate different quantization
axes, due to 120$^\circ$ misalignment of local moments. Thus, the hopping electron is always in a superposition of spin-up and spin-down
states. }
\label{fig_cube_2}
\end{figure}

The transport properties of the system are described within the
nearest-neighbor tight-bonding approximation validated by low
carrier density \cite{Chi2001307} and high resistivity
\cite{Chu2012173} of Mn-based antiperovskites. In our toy model we
consider only those electrons that hop between Mn sites as they
give the main contribution to the transport
properties.\cite{Jardin:1975,Takenaka:apl/98/2/10.1063/1.3541449,Ali2014141}
Then the local Hamiltonian for the conduction electrons takes a
form:
\begin{equation}
\hat{H}\left( \mathbf{r},t\right) =\sum_{j\tau }\varepsilon _{0}\left(
\mathbf{k}\right) \hat{a}_{j\tau }^{\dag }\hat{a}_{j\tau }+\sum_{j,l,\tau}\gamma
_{jl}\left( \mathbf{k}\right) \hat{a}_{j\tau }^{\dag }\hat{a}_{l\tau
}-J_{sd}\sum_{j\tau ,\tau ^{\prime }}\mathbf{S}_{j}\left( t,\mathbf{r}%
\right) \hat{a}_{j\tau }^{\dag }\hat{\boldsymbol{\sigma }}_{\tau \tau
^{\prime }}\hat{a}_{j\tau ^{\prime }},  \label{eq_hamiltonian}
\end{equation}%
where the first term in the r.h.s. describes the kinetic energy related with the
crystal translational symmetry. Fermi-operators $\hat{a}_{j\tau }$ and $\hat{%
a}_{j\tau }^{\dag }$ describe annihilation/creation of the electrons with the Bloch functions $\left\vert
u_{j}\right\rangle $ ($\left\langle u_{k}\mid u_{j}\right\rangle =\delta
_{kj}$) and in the spin states $\left\vert \tau \right\rangle $ ($\tau
=\uparrow ,\downarrow $): $\hat{a}_{j\tau }^{\dag }\left\vert 0\right\rangle
=\left\vert u_{j}\right\rangle \left\vert \tau \right\rangle $ at
different sublattices  $j=1,2,3$.
In what follows we neglect the dispersion of $\varepsilon _{0}\left( \mathbf{%
k}\right) $ and set its value to zero. Coefficient $\gamma _{jl}(\mathbf{k}%
)=-\sum_{\boldsymbol{\delta }}t_{\delta }e^{i\mathbf{k}\boldsymbol{\delta }}$
is the hopping term between neighboring sites (connected with $%
\boldsymbol{\delta }$) that belong to different sublattices $j$
and $l$ (see Fig.\ref{fig_cube_2}). Constant $J_{sd}$ describes the
exchange coupling between localized and free electrons (so-called
$sd$-exchange), it can be either
positive or negative, and without loss of generality we take $J_{sd}>0$. $%
\hat{\boldsymbol{\sigma }}$ is the spin operator.

The local band structure obtained by diagonalization of Hamiltonian (\ref{eq_hamiltonian}) consists of six bands which, neglecting the $sd$%
-exchange, are pair-wise (spin-up and spin-down) degenerate. 
The exchange
interaction gives rise to an additional splitting and mixing of bands. As
sublattice sites 1, 2 and 3 are equivalent, $\gamma _{12}(\mathbf{k})=\gamma
_{23}(\mathbf{k})=\gamma _{31}(\mathbf{k})=\gamma (\mathbf{k})<0$.
In this case the band structure splits into four bands ($\varepsilon
_{1}<\varepsilon _{2}<\varepsilon _{3}<\varepsilon _{4}$):
\begin{eqnarray}
\varepsilon _{1}(\mathbf{k}) &=&2\gamma -J_{sd}S,\quad \varepsilon _{2}(%
\mathbf{k})=-\gamma -\sqrt{J_{sd}^{2}S^{2}+9\gamma ^{2}},  \nonumber
\label{eq_enegy_bands} \\
\varepsilon _{3}(\mathbf{k}) &=&2\gamma +J_{sd}S,\quad \varepsilon _{4}(%
\mathbf{k})=-\gamma +\sqrt{J_{sd}^{2}S^{2}+9\gamma ^{2}}.
\end{eqnarray}%
The states in the bands $\varepsilon _{1},\varepsilon _{3}$ are nondegenerate,
and those in bands $\varepsilon _{2},\varepsilon _{4}$ are
double-degenerate. Corresponding eigen vectors (local spin quantization axis
is parallel to $\mathbf{L}_{1}$) for the lower bands $\varepsilon
_{1},\varepsilon _{2}$ are the following:
\begin{eqnarray}
\left\vert \Psi _{1}\right\rangle &=&\frac{1}{\sqrt{3}}\left\vert \uparrow
\right\rangle \left[ \left\vert u_{1}\right\rangle +\frac{1}{2}\left(
\left\vert u_{2}\right\rangle -\left\vert u_{3}\right\rangle \right) \right]
+\frac{1}{2}\left\vert \downarrow \right\rangle \left( \left\vert
u_{2}\right\rangle +\left\vert u_{3}\right\rangle \right) ,  \nonumber
\label{eq_eigen_states} \\
\left\vert \Psi _{2a}\right\rangle &=&\left\vert \eta _{1}\right\rangle
\left\vert \uparrow \right\rangle +\left\vert \eta _{2}\right\rangle
\left\vert \downarrow \right\rangle ,\quad \left\vert \Psi
_{2b}\right\rangle =\left\vert \eta _{2}\right\rangle \left\vert \uparrow
\right\rangle +\left\vert \eta _{3}\right\rangle \left\vert \downarrow
\right\rangle ,
\end{eqnarray}%
where we have introduced the following (non-normalized) combinations of the
mutually orthogonal Bloch functions (see Appendix \ref{app:diagonalization} for details):
\begin{eqnarray}
\left\vert \eta _{1}\right\rangle &=&\sqrt{\frac{2}{3}}\cos \psi \left\vert
u_{1}\right\rangle -\frac{1}{2\sqrt{6}}\left( \cos \psi +3\sin \psi \right)
\left( \left\vert u_{2}\right\rangle -\left\vert u_{3}\right\rangle \right) ,
\nonumber  \label{eq:functions_for_calculation} \\
\left\vert \eta _{2}\right\rangle &=&\frac{1}{2}\sin \left( \psi -\frac{\pi
}{4}\right) \left( \left\vert u_{2}\right\rangle +\left\vert
u_{3}\right\rangle \right) , \\
\left\vert \eta _{3}\right\rangle &=&\sqrt{\frac{2}{3}}\sin \psi \left\vert
u_{1}\right\rangle -\frac{1}{2\sqrt{6}}\left( \sin \psi +3\cos \psi \right)
\left( \left\vert u_{2}\right\rangle -\left\vert u_{3}\right\rangle \right)
\nonumber \\
\left\langle \eta _{1}\mid \eta _{1}\right\rangle &=&\left\langle \eta
_{3}\mid \eta _{3}\right\rangle =\frac{1}{4}\left( 3+\sin 2\psi \right)
,~\left\langle \eta _{2}\mid \eta _{2}\right\rangle =\frac{1}{4}\left(
1-\sin 2\psi \right) .  \nonumber
\end{eqnarray}%
The effective parameter $\psi$ depends on the relation between the $sd$-exchange
and the hopping integral as follows:
\begin{equation}
\sin 2\psi =\frac{3\gamma }{\sqrt{J_{sd}^{2}S^{2}+9\gamma ^{2}}}.
\label{eq:sinus}
\end{equation}%
It is determined by the band structure and, like in the case of
collinear AFM, \cite{Cheng:PhysRevB.86.245118} plays a crucial
role in the adiabatic electron dynamics. It describes an
overlap of the functions (\ref{eq:functions_for_calculation}),
\begin{equation}
~\left\langle \eta _{1}\mid \eta _{3}\right\rangle =\frac{1}{4}\left(
1+3\sin 2\psi \right) ,  \label{eq:overlap}
\end{equation}%
and hence, the spin tunneling between different sites.

Expressions for the eigen functions $\left\vert \Psi _{3}\right\rangle $, $%
\left\vert \Psi _{4a,b}\right\rangle $ corresponding to the upper bands $%
\varepsilon _{3},\varepsilon _{4}$, are analogous to those for
$\left\vert \Psi _{1}\right\rangle ,\left\vert \Psi
_{2a,b}\right\rangle$ with substitution $\psi \rightarrow -\psi $,
$\uparrow \leftrightarrow \downarrow $.

Rather complicated (compared to the case of collinear AFM) structure of the
eigen-functions $\left\vert \Psi _{j}\right\rangle$ is due to
noncollinearity of neighboring localized moments. As a result, a free
electron polarized along, say, $\mathbf{S}_{1}$ direction is, after hopping,
always in a superposition of spin states with respect to the new host's
quantization axis (see Fig.\ref{fig_cube_2}).
It is instructive to analyze the energy spectrum with account of average
spin $\mathbf{s\equiv }\left\langle \hat{\boldsymbol{\sigma }}\right\rangle $
of the corresponding eigen state (hereafter we use the convention $\hbar=1$%
). As it was already mentioned, in the AFM ground state the magnetization of
localized spins $\mathbf{M}=0$  and one can
anticipate that the ground state of conduction electrons is a spin-less.
It can be easily checked from (\ref{eq_eigen_states}) that the
lowest energy band ($\varepsilon _{1}$) corresponds to the zero-spin ``singlet'' state
$\left\vert \Psi _{1}\right\rangle$. 
Next band in energy
scale, $\varepsilon _{2}$, is formed by degenerate states $\left\vert \Psi
_{2a}\right\rangle $ and $\left\vert \Psi _{2b}\right\rangle $ which are
spin-polarized in $z$ direction parallel to AFM vector $\mathbf{L}_1$ with
 opposite spin values $S_z=\pm(1+\sin2\psi)/2$. In equilibrium ($\mathbf{M%
}=0$) both states should be equally populated.  The other states that form
the upper bands $\varepsilon _{3}$ and $\varepsilon _{4}$ have analogous
properties: $\left\vert \Psi _{3}\right\rangle $ is spin-less, and $%
\left\vert \Psi _{4a}\right\rangle$, $\left\vert \Psi _{4b}\right\rangle $
are spin-polarized.

Obviously, in the case of spin-injection only the degenerate states $\left\vert \Psi _{2a}\right\rangle $ , $%
\left\vert \Psi _{2b}\right\rangle $ \ and $\left\vert \Psi
_{4a}\right\rangle $, $\left\vert \Psi _{4b}\right\rangle$ could
be populated and thus can participate in spin transport. Moreover,
in the system under consideration the Berry connection
of nondegenerate states is proportional to the average spin (see Appendix %
\ref{app:Berry_connection}) and thus vanishes for $\left\vert \Psi
_{1}\right\rangle $ and $\left\vert \Psi _{3}\right\rangle $. If,
in addition, $s-d$ exchange coupling is rather strong ($J_{sd}\gg
\gamma$), the lower bands $\varepsilon_{1},\varepsilon_{2} $ are
well separated from the upper ones
$\varepsilon_{3},\varepsilon_{4}$, and the transport of
spin-polarized electrons is restricted mainly to the second $\varepsilon_{2}$ band.

In what follows we assume that the Fermi level is situated in the vicinity
of the degenerate band $\varepsilon _{2}$ and in the next section we consider the
adiabatic spin dynamics related with tunneling between states $\left\vert
\Psi _{2a}\right\rangle $ and $\left\vert \Psi _{2b}\right\rangle$ in the AFM
texture.

\subsection{Pseudospin and dynamic equations}

We follow the semiclassical approach
\cite{Cheng:PhysRevB.86.245118,Sundaram:PhysRevB.59.14915,Xiao:RevModPhys.82.1959,
Culcer:PhysRevB.72.085110} to describe the effective electron
dynamics in the degenerate band $\varepsilon _{2}$. An individual
electron is seen as a wave-packet
\begin{equation}
\left\vert W\right\rangle =\int d\mathbf{k}p\mathbf{(k-k}_{c})\left[
c_{a}\left\vert \Psi _{2a}\right\rangle +c_{b}\left\vert \Psi
_{2b}\right\rangle \right] ,  \label{eq:wave_packet}
\end{equation}
where $\int d\mathbf{k}\left\vert p\mathbf{(k-k}_{c})\right\vert ^{2}=%
\mathbf{k}_{c}$ is the center of mass momentum, $\left\langle W\right\vert
\mathbf{r}\left\vert W\right\rangle =\mathbf{r}_{c}$ is the center of mass
position, $\left\vert c_{a}\right\vert^{2}+\left\vert c_{b}\right\vert
^{2}=1$. We assume that the wave-packet (\ref{eq:wave_packet}) spreads small
compared to the length-scale of AFM inhomogenuity.

Coherent dynamics between the two subbands introduces an internal degree of
freedom which we describe by the spinor $(c_{a} c_{b})$ or, equivalently, by the isospin vector
\begin{equation}
\mathbf{C}=\left\{ 2\mathrm{Re}(c_{a}c_{b}^{\ast }),-2\mathrm{Im}%
(c_{a}c_{b}^{\ast }),\left\vert c_{a}\right\vert ^{2}-\left\vert
c_{b}\right\vert ^{2}\right\},  \label{eq:isospin}
\end{equation}
It is worth to menition that both the spinor and the normalized isospin vector $\mathbf{C}$ represent $SU(2)$ group in 2-dimensional Hilbert space formed by the state vectors $\left\vert \Psi _{2a}\right\rangle$ and $\left\vert \Psi
_{2b}\right\rangle$.

Space-time dependence of the state vectors $\left\vert \Psi _{2a}\right\rangle $%
\ and $\left\vert \Psi _{2b}\right\rangle$ stems exclusively from the rotation
of the local spin quantization axis induced by variation of the
AFM moments ($\mathbf{S}_{j}(t,\mathbf{r}_{c})$ or, equivalently, $\mathbf{L}_{1}(t,\mathbf{r}_{c}),\mathbf{L}_{2}(t,\mathbf{r}_{c}),\mathbf{n}(t,\mathbf{r}_{c})$). It should be noted that the
space dependence of Bloch functions $|u_j\rangle$ is substantial only at
the lengthscale of interatomic distances and thus is unimportant at the
large-scale variations of the AFM order parameters. Since, in addition, we neglect
spin-orbit interactions, rotation of the magnetic moments is decoupled from
variation of the crystallographic axes and, correspondingly, spin and
space-dependent states of the carriers are disentangled.

As the local orientation of the AFM moments is unambiguously defined by
the rotation matrix $\Re (\boldsymbol{\varphi )}$ (see (\ref{eq:Gibbs_vector}%
)), the state vector $\left\vert \Psi _{2a,b}(r_{\mu })\right\rangle$ at a
given point $r_{\mu }=(t,\mathbf{r}_{c})$ can be defined by a $SU(2)$ gauge
unitary transformation corresponding to $O$(3) rotation:
\begin{equation}
\hat{U}=\cos \frac{\theta }{2}\hat{1}-i\sin \frac{\theta }{2}\mathbf{N}\hat{%
\boldsymbol{\sigma }}=\frac{1}{\sqrt{1+\boldsymbol{\varphi }^{2}}}\left(
\hat{1}-i\boldsymbol{\varphi }\hat{\boldsymbol{\sigma }}\right).
\label{eq:unitary transformation}
\end{equation}

Thus,
\begin{equation}
\left\vert \Psi _{2}(r_{\mu })\right\rangle =\hat{U}\left\vert \Psi
_{2}(r_{\mu }^{0})\right\rangle  \label{eq:transformaed state}
\end{equation}
where the reference state vectors $\left\vert \Psi _{2}(r_{\mu
}^{0})\right\rangle$ and reference AFM vectors $\mathbf{L}_{1}^{(0)},
\mathbf{L}_{2}^{(0)}$ are taken in the same fixed point $r_{\mu }^{0}$. The
gauge is fixed by the choice of spin eigenstates at this point\footnote{~The gauge fields in non-Abelian theory are gauge covariant, not gauge invariant. However, as it was explained in details in Ref.~\onlinecite{Cheng:PhysRevB.86.245118}, ultimate quasiclassical equaitons (\ref{eq_dynamics initial_2}), (\ref{eq_dynamics initial_3}) include only the isospin scalars $\mathbf{C}\mathbf{R}_{\mu\nu}$ which respect the gauge invariance. Thus, for the sake of definetness we can fix the gauge in a similar way as in Ref.~\onlinecite{Cheng:PhysRevB.86.245118} without loss of generality.}.

The vectors $\left\vert \Psi _{2a}\right\rangle $\ and $\left\vert \Psi
_{2b}\right\rangle $ also depend indirectly on quasi-wave-vector $k_{\mu
}=(0,\mathbf{k}_{c})$ through the coefficient $\psi \left[ \gamma (\mathbf{k}%
_{c})\right]$.

According to the general theory,\cite{Xiao:RevModPhys.82.1959} the
set of equations of motion for the dynamic variables $\mathbf{r}_{c}$,
$\mathbf{k}_{c}$
and $\mathbf{C}$ can be written as follows (see Ref.~\onlinecite%
{Cheng:PhysRevB.86.245118} for the detailed derivation):
\begin{eqnarray}
\mathbf{\dot{C}} &=&2\mathbf{C\times }\left( \mathbf{A}_{\mu }^{r}\dot{r}%
_{\mu }+\mathbf{A}_{\mu }^{k}\dot{k}_{\mu }\right),
\label{eq_dynamics initial} \\
\dot{k}_{\mu } &=&-\partial _{\mu }^{r}\varepsilon_2 +\mathbf{C}\left( \mathbf{%
R}_{\mu \nu }^{rr}\dot{r}_{\nu }+\mathbf{R}_{\mu \nu }^{rk}\dot{k}_{\nu
}\right),  \label{eq_dynamics initial_2} \\
\dot{r}_{\mu } &=&\partial _{\mu }^{k}\varepsilon_2 -\mathbf{C}\left( \mathbf{R%
}_{\mu \nu }^{kr}\dot{r}_{\nu }+\mathbf{R}_{\mu \nu }^{kk}\dot{k}_{\nu
}\right),  \label{eq_dynamics initial_3}
\end{eqnarray}%
where the gauge potentials $\left\{ \mathbf{A}_{\mu }^{r},\mathbf{A}_{\mu
}^{k}\right\} $, Berry connection $\hat{\mathcal{A}}_{\mu }^{\alpha }\equiv
\mathbf{A}_{\mu }^{\alpha }\hat{\boldsymbol{\sigma }}$, and Berry curvatures
$\mathbf{R}_{\mu \nu }^{\alpha \beta }$ ($\alpha ,\beta =r,k$ ) are
introduced in a standard way as
\begin{eqnarray}
\hat{\mathcal{A}}_{\mu }^{\alpha } &=&i\left(
\begin{array}{cc}
\left\langle \Psi _{2a}\mid \partial _{\mu }^{\alpha }\Psi _{2a}\right\rangle
& \left\langle \Psi _{2a}\mid \partial _{\mu }^{\alpha }\Psi
_{2b}\right\rangle \\
\left\langle \Psi _{2b}\mid \partial _{\mu }^{\alpha }\Psi _{2a}\right\rangle
& \left\langle \Psi _{2b}\mid \partial _{\mu }^{\alpha }\Psi
_{2b}\right\rangle%
\end{array}%
\right) ,  \label{eq:gauge_potential_def} \\
\mathbf{R}_{\mu \nu }^{\alpha \beta } &=&\partial _{\mu }^{\alpha }\mathbf{A}%
_{\nu }^{\beta }-\partial _{\nu }^{\beta }\mathbf{A}_{\mu }^{\alpha }+2%
\mathbf{A}_{\mu }^{\alpha }\times \mathbf{A}_{\nu }^{\beta }.
\label{eq:Berry_curvatures}
\end{eqnarray}%
Starting from equation (\ref{eq_dynamics initial}) we drop the subscript $c$
on $\mathbf{r}_{c}$ and $\mathbf{k}_{c}$.

\section{Adiabatic dynamics}

\subsection{Berry curvature and topology of AFM texture}

Before considering the possible dynamics of free electrons it is
instructive to analyze the explicit expressions for the Berry connection and the Berry
curvature in AFM texture. Calculations based on definitions (\ref%
{eq:gauge_potential_def}), (\ref{eq:Berry_curvatures}) (see Appendix \ref%
{app:Berry_connection}) show that the gauge potential
\begin{equation}
\mathbf{A}_{\mu }^{r}=\frac{1}{4}\Re ^{-1}\left( \boldsymbol{\varphi }%
\right) \left[ \left( 1+\sin 2\psi \right) \boldsymbol{\Omega }_{\mu
}-\left( 1-\sin 2\psi \right) \mathbf{n}\left( \boldsymbol{\Omega }_{\mu }%
\mathbf{n}\right) \right] ,  \label{eq:Berry_connection}
\end{equation}%
and $\mathbf{A}_{\mu }^{k}=0$ in the absence of spin-orbit
interactions.\cite{Cheng:PhysRevB.86.245118} Correspondingly,
nontrivial components of the Berry curvature are the following
\begin{eqnarray}
\mathbf{R}_{\mu \nu }^{rr} &=&\frac{1}{8}\left( 1-\sin 2\psi \right) \Re
^{-1}\left( \boldsymbol{\varphi }\right) \left[ 2\left( 1+\sin 2\psi \right)
\mathbf{\Omega }_{\nu }\times \boldsymbol{\Omega }_{\mu }-\left( 3+\sin
2\psi \right) \mathbf{n}\left( \mathbf{n\cdot \Omega }_{\nu }\times
\boldsymbol{\Omega }_{\mu }\right) \right] ,  \nonumber
\label{eq:Berry_curvatures_2} \\
\mathbf{R}_{\mu \nu }^{rk} &=&-\mathbf{R}_{\nu \mu }^{kr}=\frac{1}{4}%
\partial _{\nu }^{k}\sin 2\psi \Re ^{-1}\left( \boldsymbol{\varphi }\right) %
\left[ \boldsymbol{\Omega }_{\mu }+\mathbf{n}\left( \boldsymbol{\Omega }%
_{\mu }\mathbf{n}\right) \right] .
\end{eqnarray}%
The multiplier $\Re ^{-1}\left( \boldsymbol{\varphi }\right) $ (inverse
rotation matrix) in (\ref{eq:Berry_connection}), (\ref{eq:Berry_curvatures_2}%
) results from the gauge covariance of the non-Abelian gauge fields. As we will
see later, the same rotation relates the isospin to the real spin.

\begin{figure}[htbp]
\centering
\includegraphics[width=0.5\linewidth]{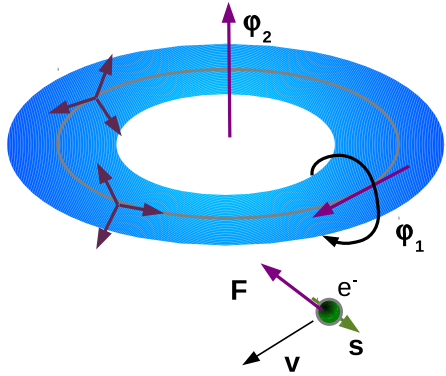}
\caption{(Color online) Orbital dynamics of an electron in the AFM
texture with nonzero curvature. Blue (shadowed) torus shows the
area with twisted (vector {$\boldsymbol{\varphi}_{1}$}) and bended
(vector {$\boldsymbol{\varphi}_{2}$}) AFM vectors (three bounded
arrows). Free spin-polarized (spin $\mathbf{s}$) electron moving
in this area with the group velocity $\mathbf{v}$ feels the
effective force $\mathbf{F}$, see text for
details.}\label{fig_cube_6}
\end{figure}

Analysis of the relations (\ref{eq:Berry_connection}), (\ref%
{eq:Berry_curvatures_2}) shows that the gauge potentials produced by the AFM texture
depend on the orientation of AFM vectors through the rotation vector $%
\boldsymbol{\Omega }_{\mu }$ which, in turn, is related to the dynamic magnetization $\mathbf{M}_{\mathrm{dyn}}$. The Berry curvature $\mathbf{R}_{\mu \nu
}^{rr} $ for \emph{free electron} is proportional to the curvature $\mathbf{K%
}_{\mu \nu }\equiv \boldsymbol{\Omega }_{\mu }\times \boldsymbol{\Omega }%
_{\nu }=\partial _{\mu }^{r}\boldsymbol{\Omega }_{\nu }-\partial _{\nu }^{r}%
\mathbf{\Omega }_{\mu }$ of the \emph{AFM texture}\footnote{%
~Note, that we discuss only the rotations in spin state, the lattice
itself is supposed to be unchanged. However, the theory can be
generalized to include space rotations of the lattice, as it will be
discussed below. See also
Ref.\onlinecite{Sundaram:PhysRevB.59.14915} for description of the
gauge fields in deformed crystals.}. 
In other words, the Berry
curvature is intimately related with topological properties of the 
space distribution of the localized moments. To give an example 
distribution with the nontrivial curvature, we note that in
elasticity theory $K_{\mu \nu }\neq 0$ is called bend-twist
tensor. So, one
can imagine that in some toroidal area (see Figs.\ref{fig_cube_6},\ref%
{fig_cube_9}) the orientation of AFM vectors is obtained by two rotations --
one, around the in-plane axis tangent to the torus (twist with the rotation vector $%
\boldsymbol{\varphi}_{1}$), and another, around the vertical torus axis $z$ (bend with the
rotation vector $\boldsymbol{\varphi}_{2}$). The curvature vector $\mathbf{K}%
_{xy}=\boldsymbol{\Omega }_{x}\times \boldsymbol{\Omega }_{y}$ in
each point of the structure is directed along the radius. This
structure, through $sd$-exchange, forms a potential with a Berry curvature $\mathbf{R}%
_{xy}^{rr}\propto \mathbf{K}_{xy}$  and
produces a ``Lorentz-like'' effective force for free spin-polarized electrons, as will be explained below.

It is instructive to compare Expr. (\ref{eq:Berry_connection}) describing a gauge
potential in AFM with the analogous expression \cite%
{Tserkovnyak:PhysRevB.77.134407} for a ferromagnetic material with
the magnetization vector $\mathbf{m}$:
\begin{equation}
\mathbf{A}_{\mu }^{FM}=-\mathbf{m}\times \partial _{\mu }^{r}\mathbf{m}=-%
\left( 1-\mathbf{m\otimes m}\right) \boldsymbol{\Omega}_{\mu}.
\label{eq:FM_gauge}
\end{equation}%
Although in ferromagnets the gauge transformation is applied to the real spin
(not to isospin), and the gauge potential is Abelian, comparison of expressions (%
\ref{eq:FM_gauge}) and (\ref{eq:Berry_connection}) shows that up to the
details related with peculariaties of electron hopping, geometric effects in
FM and AFM are related with the `` dynamic'' magnetization proportional to the
angular velocity $\boldsymbol{\Omega}_{\mu}$.


It is also worth to note the relation between the curvature $\mathbf{K}%
_{\mu \nu }$ and the topological properties of localized nonlinear
magnetic structures: skyrmions, solitons, vortecies. For example,
the topological charge of a two-dimensional (in the real space)  
skyrmion, that counts how many times the vector order parameter 
$\mathbf{m}(t,\mathbf{r})$ wraps around the unit sphere, is defined
as:\cite{Nagaosa:nnano.2013.243}
\begin{equation}
Q=\frac{1}{8\pi }\int \mathbf{m}\cdot\left( \partial _{\mu }\mathbf{m}\times
\partial _{\nu }\mathbf{m-}\partial _{\nu }\mathbf{m}\times \partial _{\mu }%
\mathbf{m}\right) dx_{\mu }dx_{\nu }.  \label{eq_curvature_FM_1}
\end{equation}%
With the use of the relations (\ref{eq:Gibbs_vector}), (\ref%
{eq:time_frequency}), and (\ref{eq:space_frequency}) it can be easily shown
that $Q$ is proportional to the projection of the curvature vector $\mathbf{K}%
_{\mu \nu }$ onto the order parameter $\mathbf{m}$ averaged over space:
\begin{equation}
Q=\frac{1}{4\pi }\int (\mathbf{m}\cdot \boldsymbol{\Omega }_{\mu }\times
\boldsymbol{\Omega }_{\nu })dx_{\mu }dx_{\nu }.  \label{eq_curvature_FM_2}
\end{equation}
The relation (\ref{eq_curvature_FM_2}) is applicable to any magnetic system with the vector order parameter: a ferromagnet, with $\mathbf{m}$ playing the role of magnetization vector, a collinear AFM, where  $\mathbf{m}$ corresponds to the N\'eel vector.  

Topological charges of the three-dimensional structures with the vector order parameter are characterized with the Hopf's invariant \cite{Volovik:0022-3719-20-7-003, Ivanov:1983E} which describes  $S^3$ to $S^2$ map:
\begin{equation}\label{eq_Hopf_invariant}
H=\frac{1}{16\pi^2}\int dx^3\varepsilon_{\mu\nu\gamma}\varepsilon_{jklm}\nu_j\frac{\partial \nu_k}{\partial x_\mu}\frac{\partial\nu_l}{\partial x_\nu}\frac{\partial\nu_m}{\partial x_\gamma},
\end{equation}
where the four-component vector $(\boldsymbol{\nu}, \nu_4)$ is related with the Gibb's vector $\boldsymbol{\varphi}=\boldsymbol{\nu}/\nu_4$, $\nu_4=\cos\theta/2$, $\varepsilon_{\mu\nu\gamma}$ and $\varepsilon_{jklm}$ are fully antisymmetric tensors.

For the case of noncollinear AFMs an appropriate topological invariant is given by the expressions \cite{Rajaraman}
\begin{equation}\label{eq_topologica_invariant_S3_s3}
Q=-\frac{1}{24\pi^2}\int dx^3\varepsilon_{\mu\nu\gamma}\mathrm{Tr}\left[\Re^{-1}(\boldsymbol{\varphi})\frac{\partial }{\partial x_\mu}\Re(\boldsymbol{\varphi})\Re^{-1}\frac{\partial }{\partial x_\nu}\Re(\boldsymbol{\varphi})\Re^{-1}\frac{\partial }{\partial x_\gamma}\Re(\boldsymbol{\varphi})\right],
\end{equation}
which corresponds to  $S^3$ to $S^3$ map. Three-dimensional AFM order parameter (formed by mutually orthogonal $\mathbf{L}_1$ and $\mathbf{L}_2$ vectors) is parametrized with the rotational tensor $\Re(\boldsymbol{\varphi})$. With the use of expressions (\ref{eq:time_frequency}) the  topological invariant (\ref{eq_topologica_invariant_S3_s3}) can be expressed in terms of the rotation vectors $\boldsymbol{\Omega}_\mu$ as follows:
\begin{equation}\label{eq_topologica_invariant_omega}
Q=-\frac{1}{24\pi^2}\int dx^3\varepsilon_{\mu\nu\gamma}\left(\boldsymbol{\Omega}_\mu\cdot\boldsymbol{\Omega}_\nu\times\boldsymbol{\Omega}_\gamma\right).
\end{equation}
It can be easily seen that, in analogy with ferromagnets, the topological charge (\ref{eq_topologica_invariant_omega}) is proportional to the projection of the curvature vector $\mathbf{K}
_{\nu\gamma }$ onto diraction of dynamics magnetization $\boldsymbol{\Omega}_\mu$ seen by the free electron moving in $x_\mu$ direction.

Note that the model ``twist-bend'' structure shown in Figs.\ref{fig_cube_6},\ref{fig_cube_9} can, in principle, have nonzero topological charge  if the the AFM ordering outside the toroidal area is homogeneous. In this case the structure is characterized with three nontrivial noncollienar rotational vectors:  $\boldsymbol{\Omega}_1$ and $\boldsymbol{\Omega}_2$ for twisting and bending and $\boldsymbol{\Omega}_3$ that describes smooth rotation of AFM vectors in the intermediate area between torus and infinity.

\subsection{Free spin dynamics and gauge potential}

Let us consider a typical spintronic problem in which a nonequilibrium
spin-polarized carrier is injected into an AFM. The question is: ``Does the AFM
medium affects the state of an electron? If it does, how could this effect be
observed?''

\begin{figure}[tbp]
\centering
\includegraphics[width=0.5\linewidth]{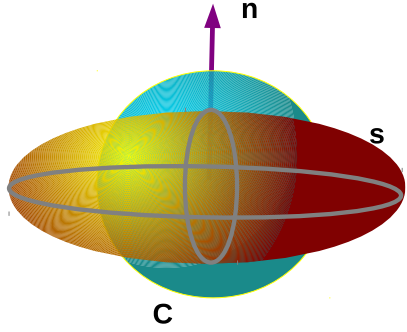}
\caption{(Color online) Isospin $\mathbf{C}$ (sphere) and real
spin $\mathbf{s}$ (spheroid) in the local frame.}
\label{fig_cube_5}
\end{figure}

To begin with, we find the relation between the isospin vector $\mathbf{C}$
(which itself is not gauge invariant and depends upon the choice of the eigen
functions $\left\vert \Psi _{2a}\right\rangle $, $\left\vert \Psi
_{2b}\right\rangle $) and the observable (and fully gauge invariant) spin $\mathbf{s\equiv }%
\left\langle W\right\vert \hat{\boldsymbol{\sigma }}$ $\left\vert
W\right\rangle $. Direct calculations based on the states (\ref%
{eq_eigen_states}), (\ref{eq:transformaed state}) give rise to the following
expression for spin vector:
\begin{equation}
\mathbf{s}(t,\mathbf{r})=\frac{1}{2}\left( 1+\sin 2\psi \right) \left[ \Re
\left( \boldsymbol{\varphi }\right) \mathbf{C}\right] -\frac{1}{2}\left(
1-\sin 2\psi \right) \left( \Re \left( \boldsymbol{\varphi }\right) \mathbf{%
C\cdot n}\right) \mathbf{n}.  \label{eq:spin_isospin}
\end{equation}%
Using normalization of the isospin, $\left\vert \mathbf{C}\right\vert ^{2}=1$,
we arrive at a relation between the spin of free carrier and
the orientation of the plane formed by the localized spins and represented by the
vector $\mathbf{n(}t,\mathbf{r)}$ (see Fig.\ref{fig_cube_5}):
\begin{equation}
\frac{\left( \mathbf{ns}\right) ^{2}}{\sin ^{2}2\psi }+\frac{4\left( \mathbf{%
n\times s}\right) ^{2}}{\left( 1+\sin 2\psi \right) ^{2}}=1.
\label{eq:spin_normal relation}
\end{equation}

Analysis of expression (\ref{eq:spin_normal relation}) shows that, like
in the case of a collinear AFM \cite{Cheng:PhysRevB.86.245118} (see also Eq.(%
\ref{eq:Cheng_8a})), spin polarizarion of the conduction electron depends
upon the orientation of the AFM vectors. In a noncollinear AFM the vector $\mathbf{s}$
varies on an oblate spheroid (on a prolate in the collinear AFM), the short
axis of which coincides with the plane normal. Like in the collinear case, $%
\mathbf{s}^{2}\leq 1$, which means that an electron is in a mixed spin state
due to entanglement between the space and spin degrees of freedom.

Equation for spin dynamics obtained from (\ref{eq:spin_isospin}) with
account of (\ref{eq_dynamics initial}) and (\ref{eq:Berry_connection}) (see
Appendix \ref{app:Berry_connection} for hints of the derivation) is similar
to the Euler's equation for rotation of a rigid body:
\begin{equation}
\dot{\mathbf{s}}-\boldsymbol{\Omega }\times \mathbf{s}=-\sin 2\psi
\boldsymbol{\Omega }\times \hat{g}\mathbf{s},  \label{eq:motions_1}
\end{equation}%
where we introduced the tensor
\begin{equation}
\hat{g}=\hat{1}\mathbf{+}\left[ \left( \frac{1+\sin 2\psi }{2\sin 2\psi }%
\right) ^{2}-1\right] \mathbf{n\otimes n},  \label{eq:effective-field_1}
\end{equation}%
and, as it was already noted, $\boldsymbol{\Omega }=\boldsymbol{\Omega }%
_{t}+\boldsymbol{\Omega }_{l}\dot{r}_{l}$. The second term on the
l.h.s. of Eq. (\ref{eq:motions_1}) originates from the rotation of the local
frame associated with vectors $\mathbf{L}_{1}$, $\mathbf{L}_{2}$, $\mathbf{n}
$ (for the constant isospin $\mathbf{C}$). In analogy with Ref.\onlinecite%
{Sundaram:PhysRevB.59.14915}, it could be called ``tracking'' term  because it reflects the tendency of the AFM
lattice to drag the electron spin along with the time/space AFM motion.

The r.h.s. describes the spin evolution in the local frame due to
accumulation of a $SU$(2) non-Abelian Berry phase. Namely, spin vector
rotates around the angular velocity $\boldsymbol{\Omega }$ being
simultaneously bounded to the spheroid (\ref{eq:spin_normal relation}).
Thus, Eq.~(\ref{eq:motions_1}) is the Bloch equation for spin precession in
the effective magnetic field $\mathbf{H}^{eff}=-\sin 2\psi \hat{g}%
\boldsymbol{\Omega }$. As $\boldsymbol{\Omega }$ is proportional
to the dynamic magnetization $\mathbf{M}_{\mathrm{dyn}}$ of he AFM
layer, the origin of the effective magnetic field
$\mathbf{H}^{eff}$ acting on free spins has the same nature as in
ferromagnets.\cite{Bazaliy:PhysRevB.57.R3213}

It should be stressed that spin dynamics substantially depends upon the
strength of $sd$-exchange coupling. In the limit $J_{sd}\rightarrow 0$ (which means that $%
\sin2\psi\rightarrow 1$), the electron spin coincides with the pseudospin $%
\mathbf{C}$ up to rotation $\Re(\boldsymbol{\varphi})$ and $\dot{\mathbf{s}}%
=0$. In the opposite case of extremely strong $J_{sd}\rightarrow \infty$ (or, equvalently, $%
\sin2\psi\rightarrow 0$) the r.h.s. term in equation (\ref{eq:motions_1})
vanishes. So, in the case of strong coupling between free and localized
spins the free spin simply tracks orientation of the local frame in each point.

Orbital dynamics of the wave-packet (\ref{eq:wave_packet}) is described by
semiclassical equations obtained from (\ref{eq_dynamics initial_2}), (\ref{eq_dynamics initial_3}) with account of (\ref{eq:motions_1})\ as follows:
\begin{eqnarray}
\dot{r}_{\mu } &=&\partial _{\mu }^{k}\varepsilon_2 +\frac{1}{2}\partial _{\mu
}^{k}\ln \left( 1+\sin 2\psi \right) \left[ \mathbf{s\cdot \mathbf{\Omega }+}%
\frac{1}{\sin 2\psi }\left( \mathbf{n\cdot s}\right) \left( \mathbf{n\cdot
\Omega }\right) \right] ,  \label{eq:motions_2} \\
\dot{k}_{\mu } &=&-\partial _{\mu }^{r}\varepsilon_2 -\frac{1}{2}\mathbf{\dot{s%
}\cdot \mathbf{\Omega }}_{\mu }.  \label{eq:motions_4}
\end{eqnarray}

The spin-dependent addition to the group velocity in Eq.
(\ref{eq:motions_2}) is proportional to $\partial _{\mu
}^{k}\varepsilon_2 $, because $\partial _{\mu }^{k}\sin 2\psi \sim
\partial _{\mu }^{k}\gamma \sim \partial _{\mu }^{k}\varepsilon_2 $.
So, coupling between the free and the localized spins in the rotated AFM
texture results in ``renormalization'' of the effective electron
mass. An analogous term, omitted in (\ref{eq:motions_4}) for the sake
of simplicity, appears also in the equation for acceleration (see
Appendix \ref{app:Berry_connection},
Eq.(\ref{eq:motion_acceleration_full})).

The nontrivial, spin-dependent term on the r.h.s of Eq.
(\ref{eq:motions_4}) is intimately related to the spin dynamics. It
can also be represented in the form of a fictious Lorentz force
with the effective electric-like, $E_{\mu }$, and magnetic-like,
$B_{\xi }$ components:
\begin{eqnarray}
\dot{k}_{\mu } &=&-\partial _{\mu }^{r}\varepsilon_2 +q\left( E_{\mu
}+\varepsilon _{\mu \nu \xi }\dot{r}_{\nu }B_{\xi }\right) ,
\label{eq:electric_field} \\
qE_{\mu } &=&\frac{1}{2}\left[ \mathbf{s\cdot \mathbf{\Omega }}_{t}\mathbf{%
\times \mathbf{\Omega }}_{\mu }\mathbf{+}\sin 2\psi \hat{g}\mathbf{s\cdot
\mathbf{\Omega }}_{t}\mathbf{\times \mathbf{\Omega }}_{\mu }\right] ,
\label{eq:Lorentz_force} \\
qB_{\xi } &=&\frac{1}{2}\varepsilon _{\xi \mu \nu }\left[ \mathbf{s\cdot
\mathbf{\Omega }_{\nu }\times \mathbf{\Omega }_{\mu }+}\sin 2\psi \hat{g}%
\mathbf{s\cdot \mathbf{\Omega }_{\nu }\times \mathbf{\Omega }_{\mu }}\right]
,  \label{eq:magnetic_field}
\end{eqnarray}%
where, as above,  $\varepsilon _{\xi \mu \nu }$ is the antisymmetric Levi-Civita tensor.
The corresponding effective ``charge'' $q$
(field ``source'') is proportional to the
spin\footnote{%
Strictly speaking, an effective charge should be introduced as a vector
quantity, so, for the sake of simplicity we introduce a combination of charge
and fields. However, Faraday's relation for fields {$\varepsilon _{\xi \mu
\nu }\partial _{\mu }E_{\nu }+\partial _{t}B_{\xi }=0$} is satisfied.}.

Equations (\ref{eq:electric_field}), (\ref{eq:Lorentz_force}), and
(\ref{eq:magnetic_field}) are similar to equations that describe an
orbital motion of individual electron in a collinear
AFM.\cite{Cheng:PhysRevB.86.245118} In both cases the gauge charge
$q$ depends upon the spin which, according to equation
(\ref{eq:motions_1}), shows its own dynamics and thus can vary
in time. In both cases the gauge charge depends upon the $sd$-exchange constant (in our case, through the multiplier $\sin 2\psi
$) and vanishes when $J_{sd}\rightarrow 0$.

The new feature demonstrated in this paper is certain universality of spin-dependent
orbital dynamics in AFMs. Really, the dynamic equations of a collinear AFM ( Eqs.%
(8) of Ref.\onlinecite{Cheng:PhysRevB.86.245118}) have practically
the
same form as equations (\ref{eq:motions_1}), (\ref{eq:motions_2}), and (\ref%
{eq:motions_4}) of the present paper when written in terms of angular
velocity $\mathbf{\dot{L}=\Omega \times L}$:
\begin{eqnarray}
\dot{\mathbf{s}}-\boldsymbol{\Omega }\times \mathbf{s} &=&-\boldsymbol{%
\Omega }\times \hat{g}\boldsymbol{\mathbf{s}},\boldsymbol{~}\hat{g}=\hat{1}%
+\left( \xi ^{2}-1\right) \mathbf{L\otimes L,}  \label{eq:Cheng_8a} \\
\dot{r}_{\mu } &=&-\partial _{\mu }^{k}\varepsilon_2 +\frac{1}{2}\partial
_{\mu }^{k}\ln \xi \left[ \mathbf{s\cdot \boldsymbol{\Omega -}}\left(
\mathbf{L\cdot s}\right) \left( \mathbf{L\cdot \Omega }\right) \right] ,
\label{eq:Cheng_8b} \\
\dot{k}_{\mu } &=&-\frac{1}{2}\boldsymbol{\Omega }_{\mu }\mathbf{\cdot }\dot{%
\mathbf{s}},  \label{eq:Cheng_8c}
\end{eqnarray}
where $\xi $, in the original notations, is the overlap of the wave
functions analogous to $\sin 2\psi $ of the present paper (see
Eq.(\ref{eq:sinus})), and notations $\mathbf{L}$ is used for the
N\'{e}el vector of collinear AFM.

Thus, in AFMs with strong exchange couping between the magnetic sublattices the
emergent Lorentz force (\ref{eq:electric_field}) is defined by the \emph{%
angular velocity} $\mathbf{\mathbf{\Omega }}_{t}$ (and hence, the dynamic
magnetization) and the \emph{AFM curvature} $\mathbf{\mathbf{\Omega }_{\nu
}\times \mathbf{\Omega }_{\mu }}$, no matter how complicated or simple
the magnetic structure is. The details of the structure (number of sublattices,
dimensionality, type, mutual orientation) reveal themselves in a cumbersome
presentation of the gauge charge, tensor $\hat{g}$ and spin ellipsoid: in
a collinear AFM the anisotropy of these parameters is dictated by the easy-axis
parallel to a single AFM vector, in the present case -- by the easy-plane formed
by two orthogonal N\'{e}el vectors.

The effective electric field (\ref{eq:Lorentz_force}) plays a role of
a spin-dependent motive force (similar to that in FM, see Ref.~\onlinecite%
{Barnes:PhysRevLett.98.246601}). So, an oscillating ($\mathbf{\mathbf{\Omega
}}_{t}\neq 0$) inhomogeneous ($\mathbf{\mathbf{\Omega }}_{\mu }\neq 0$) AFM
structure can produce an electric current (or voltage) and thus can be
observed by standard electric measurements.

The effective magnetic field (\ref{eq:magnetic_field}) is proportional to the projection of
curvature $\boldsymbol{\Omega }_{\nu }\times\boldsymbol{\Omega }%
_{\mu }$ of the AFM texture on the free spin $\mathbf{s}$. Direction of the latter correlates (and in the limit of strong $sd$-exchange couling, coincides) with the direction of $\boldsymbol{\Omega}$ and dynamic magnetization of the AFM. Thus, the flux of the emergent magnetic field is
related with the topological charge of the AFMs distribution (see Eq. (\ref{eq_topologica_invariant_omega}) where the curvature is projected on the direction of dynamic magnetization seen by the free electron). An analogous situation takes place in skyrmions where the flux of emergent magnetic field is associated
with skyrmion number (i.e., topological
charge).\cite{Nagaosa:nnano.2013.243} However, gauge effects in
skyrmions and AFMs are principally different. Namely, a skyrmion can
be treated as a ferromagnet with a complicated distribution of
localized moments, and in adiabatic limit the spin of a conduction
electron always tracks the local orientation of magnetization;
gauge potential is Abelian; topological charge of skyrmion is
related with chirality.\cite{Christoph_Surgers:2014} In contrast, the
AFM which we consider here has zero total static magnetization;
complicated slowly varying magentic texture is locally formed by
a proper number of AFM vectors attributed to same point;
a conduction electron feels the local frame induced by magnetic
moments but not the direction of magnetization itself. Nontrivial
effect of
geometric phase in absence of total magnetization is due to the \emph{%
degeneracy of states} and resulting \emph{non-Abelian} character of the
gauge potential.

The emergent magnetic field in adiabatic motion is usually associated
with topological Hall effect. On the other hand, most AFM
structures are usually invariant with respect to time reversal
symmetry and this forces Hall conductivities to vanish. Recently,
it was demonstrated that
noncollinear \cite{MacDonald:PhysRevLett.112.017205} and frustrated \cite%
{Shindou:PhysRevLett.87.116801} AFMs can show anomalous Hall
effect that arises due to spin-orbit coupling. The origin of
nonzero Hall conductivity in these materials is the \emph{Berry
curvature} in \emph{momentum} space. In contrast, equation
(\ref{eq:electric_field}) (and the analogous equation for collinear
AFM from Ref.`\onlinecite{Cheng:PhysRevB.86.245118}) shows that
Hall effect can in principle be observed in compensated AFMs with
negligible spin-orbit coupling. In this case the Hall conductivity
originates from the \emph{Berry curvature} in \emph{real}
space and, what is also important, is related with the
\emph{curvature of the AFM texture}. The required breaking of time-reversal
symmetry results from
the inhomogeneuity\footnote{%
Formally, it is the combination $q\mathbf{B}$ that has pseudo-vector
character, due to the dependence on pseudovector $\mathbf{s}$ (see
(\ref{eq:magnetic_field})). So, in our case time-inversion
symmetry is broken rather by the effective charge than by the
effective field.} and related dynamic magnetization
$\mathbf{M}_{\mathrm{dyn}}\sim \boldsymbol{\Omega }$ of AFM structure as
was explained above. Following
Ref.~\onlinecite{Bruno:PhysRevLett.93.096806} this Hall effect
should be called topological.

Equations (\ref{eq:motions_1}), (\ref{eq:motions_2}), and (\ref{eq:motions_4}) state the main result of this paper. They describe electron dynamics in
a noncollinear AFM in terms of three observables ($\mathbf{s,r,k}$). However,
so far our treatment was quite general and abstract. To make physical
meaning of the obtained results clearer, in the next Section I consider
some special cases accessible for the experimental implementation.

\section{Examples}

\subsection{Travelling between AFM domains}

\begin{figure}[tbp]
\centering
\includegraphics[width=0.5\linewidth]{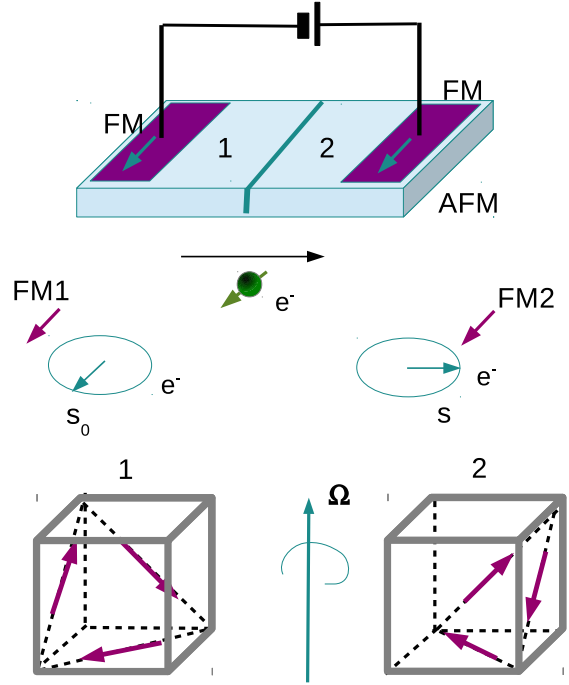}
\caption{(Color online) Probing the domain wall with the spin-polarized
current. Two ferromagnetic electrodes, FM1 and FM2, placed over the AFM film produce
spin-polarized current in the in-plane geometry (upper panel). Two
AFM domains 1 and 2 (lower panel) are related through the 90$^circ$
rotation. The spin of the electron e$^{-}$ travelling through the
domain wall tracks the local AFM structure and rotates (central
panel). Misorientation between the free spin and the FM2 magnetization
contributes to magnetoresistance and can be detected.}
\label{fig_cube_7}
\end{figure}

Let us start from the ``canonical'' example of an AFM texture -- flat,
one-dimensional domain wall separating two domains with different
orientation of spin-ordering plane (see Fig.\ref{fig_cube_7}).

For the definiteness I consider domains related by a rotation around the
cubic axis through 90$^\circ$ (corresponding rotation matrix is $\Re _{12}$).
Let us suppose that AFM film is connected to two ferromagnetic electrodes that can produce
a spin-polarized current (with the spin vector $\mathbf{s}_{0}$) in the in-plane
geometry. The thickness of the AFM layer is smaller or comparable with the
spin-coherence length (to exclude spin scattering processes), the thickness
of the domain wall is much larger than the lattice constant (to validate
the adiabatic approximation). Electric voltage applied to the system initiates
an electron flow between the ferromagnetic electrodes with an average constant velocity $%
\mathbf{v}$ directed along the domain wall normal (denoted as $x$ axis). As
the curvature of the flat domain wall is zero, the AFM texture affects only the spin
evolution described by Eq.~(\ref{eq:motions_1}) which can be rewritten in
the following form:
\begin{equation}
\frac{d\mathbf{s}}{dx}=\boldsymbol{\Omega }_{x}\times \left( \mathbf{s}-\sin
2\psi \hat{g}\mathbf{s}\right) ,  \label{eq:rotation_equation_1}
\end{equation}%
where the rotation vector $\boldsymbol{\Omega }_{x}$ has a fixed direction.
It follows from Eq.(\ref{eq:rotation_equation_1}) that after travelling
through the domain wall vector $\mathbf{s}$ will change by the value $%
\Delta \mathbf{s}$ which consists of two part: rotation with the local AFM
frame (due to the first term in parenthesis) and geometric phase rotation
(the second term) over spheroid (\ref{eq:spin_normal relation}). In the case
of strong exchange coupling ($\sin 2\psi \ll 1$) the first effect dominates.
In this case%
\begin{equation}
\Delta \mathbf{s}\approx \Re _{12}s_{0},
\end{equation}%
and the spin polarization can evolve by 90$^circ$, as shown in Fig.\ref%
{fig_cube_7}. If, then, the magnetization $\mathbf{M}_2$ of the second FM2
electrode is varied by an external magnetic field (as, e.g., in experiments
\cite{Jungwirth:2012PhRvL.108a7201M}), magnetoresistance between the FM
electrodes will also vary depending on $(\Delta \mathbf{s}\cdot \mathbf{M}%
_2)$.

Thus, the AFM domain wall is a ``spin-active'' (in analogy with optically
active) medium. Spin rotation, quantum analogue of the Faraday rotation in
optics, results from the competition of two coherent spin states and could be
observed by the electrical measurements.

\subsection{Enveloping a soliton}

\begin{figure}[tbp]
\centering
\includegraphics[width=0.5\linewidth]{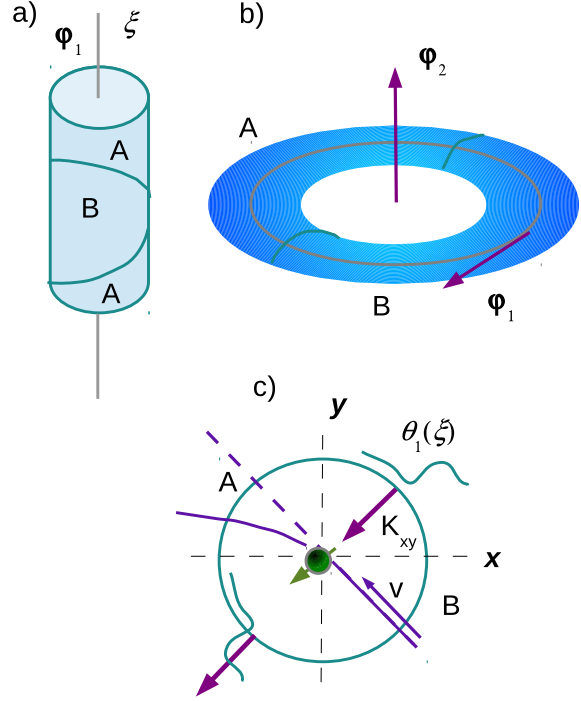}
\caption{(Color online) Electron traveling in the neighborhood of
an AFM soliton. a)
``twisting'' of an AFM structure (one dimensional rotation with the Gibb's vector $%
\boldsymbol{\varphi }_{1}$); b) ``bending'' of twisted structure; c)
spin-polarized electron moving with velocity $\mathbf{v}$ feels a
Lorentz force at the points of maximal curvature $\mathbf{K}_{xy}$ (arrows).
A and B symbolize different domains.}
\label{fig_cube_9}
\end{figure}

Equations (\ref{eq:motions_1}) and (\ref{eq:motions_4}) predict nontrivial
spin-induced orbital dynamics in a curved AFM texture. As an example, let
us consider spin-polarized electrons travelling through a region with
the inhomogeneous distribution of AFM vectors obtained in the following way.
Suppose, we start from the one-dimensional distribution of AFM vectors described
by the Gibb's vector $\boldsymbol{\varphi}_{1}=\tan \left(\theta _{1}(\xi
)/2\right) \mathbf{e}_{y}$ (a "wire" with twisted AFM structure, Fig.\ref%
{fig_cube_9}a). Next, the "wire" is bended to make a ring, this corresponds
to a rotation with Gibb's vector $\boldsymbol{\varphi }_{2}=\tan \left( \theta
_{2}(x,y)/2\right) \mathbf{e}_{z}$, where $\theta _{2}(x,y)=\tan ^{-1}y/x$
(Fig.\ref{fig_cube_9}b). The resulting rotation $\boldsymbol{\varphi
=\varphi }_{2}\circ \boldsymbol{\varphi }_{1}$ is a composition of twisting
and bending:
\begin{equation}
\boldsymbol{\varphi =\varphi }_{2}\circ \boldsymbol{\varphi }_{1}=\tan \frac{%
\theta _{1}(\xi )}{2}\mathbf{e}_{y}+\tan \frac{\theta _{2}(x,y)}{2}\mathbf{e}%
_{z}-\tan \frac{\theta _{1}(\xi )}{2}\tan \frac{\theta _{2}(x,y)}{2}\mathbf{e%
}_{x}.  \label{eq:composition_example}
\end{equation}
The effective coordinate of the first rotation, $\xi =y\cos \theta
_{2}-x\sin \theta _{2}$ bends with the "wire". Such a texture has a nonzero
curvature parallel to the radial component $\mathbf{e}_{r}$ :
\begin{equation}
\mathbf{K}_{xy}\equiv \mathbf{\Omega }_{x}\times \mathbf{\Omega }%
_{y}=-\partial _{\xi }\theta _{1}(\xi )\sin 2\theta _{2}\frac{\mathbf{e}_{r}%
}{r},  \label{eq:curvature_example}
\end{equation}%
where $r$ is the radial coordinate\footnote{%
The proposed model distribution of the AFM order parameter has a peculiarity at $%
r\rightarrow 0$ which can be avoided by placing a defect or a hole in the
center.}.

Let us further assume that the first, unbended distribution $\theta _{1}(\xi
)$ corresponds to a kink or two "head-to-head" domain walls separating
domains A and B. In this case $\partial _{\xi }\theta _{1}(\xi )\neq 0$ in
the vicinity of the domain wall localization. For the definetness we assume that the domain walls are centered at $\theta _{2}=\pi /4$ (domain wall between A and B)
and $\theta _{2}=5\pi /4$ (domain wall between A and B, see Fig.\ref{fig_cube_9}c). Obviously,   $\partial _{\xi }\theta _{1}(\xi )\neq 0$ has opposite signs
in these points. As a result, curvature $\mathbf{K}_{xy}\neq 0$ along the
line $x=y$ and has the same direction at all points.

According to Eq. (\ref{eq:magnetic_field}), AFM curvature produces fictious
out-of plane magnetic field

\begin{equation}
qB_{z}=\frac{1}{2r}\partial _{\xi }\theta _{1}(\xi )\sin 2\theta _{2}\left[
\mathbf{\mathbf{e}_{r}\cdot s+}\sin 2\psi \mathbf{e}_{r}\mathbf{\cdot }\hat{g%
}\mathbf{s}\right] .
\end{equation}

So, spin-polarized electrons with $\mathbf{s}\parallel \mathbf{K}_{xy}$
travelling with constant velocity $\mathbf{v\perp K}_{xy}$ through the
texture are exerted by the Lorentz force and deflect from the initial
trajectory\footnote{%
The spin rotation which arises during the passage through the domain wall, can
reduce the deflection. However, due to the $1/r$ dependency of the Lorentz force the
effect can still be pronounced close to the center of the region.} (see Fig.%
\ref{fig_cube_9}c).

As the curvature direction is constant, all equally polarized electrons will
deflect in the same way thusdemonstrating a topological spin Hall effect. So, such a
texture can be probed with spin-polarized current. On the other hand,
electrons with different spins are deflected in opposite directions, so, the
texture can also work as a spin-separator. It should be stressed that the
topological spin Hall effect can be modified by spin-orbit interaction neglected in
the present calculations.

\section{Conclusions}
In the present paper I consider the adiabatic dynamics of free electrons in AFM with triangular magnetic structure. I show that, in analogy with the collinear AFM,\cite{Cheng:PhysRevB.86.245118} the dynamics of a real spin $\mathbf{s}$ is influenced by space/time variation of antiferromagnetically coupled vectors $\mathbf{S}_j(\mathbf{r},t)$. The main features of the electron behavior arising from  accumulation of non-Abelian SU(2) Berry phase: precession of the electron spin around the dynamic AFM magnetization and spin-dependent orbital dynamics, --  are similar in both cases.
The first effect is of pure geometric nature, as the homogeneous AFM has no uncompensated magnetization and produces no magnetic field. The
second effect is related with processes of spin-pumping and spin-transfer
torque and thus experimentally observable.

Similarity (up to the details) of free electron dynamics  in the collinear and triangular AFMs gives the grounds to anticipate analogous effects in even more complicated AFM metals like FeMn which shows 3$\mathbf{q}$ structure described in terms of four magnetic sublattices.

The described toy model can be applied to antiperovskites
Mn$_{3}$MN (M=Ni, Ag, Zn) which show a noncollinear 120$^{\circ}$
magnetic structure and semiconducting type of conductivity (see,
e.g. Ref.\onlinecite{Lin2011452}). Although in this paper we appeal
mainly to the perovskites Mn$_3$XN, the results obtained could be
applied to the metallic AFM IrMn$_3$
 with the same triangular
structure \cite{szunyogh-2009-79,Kohn:2013} which is widely used
in spintronics due to its high N\'eel temperature.

The influence of the spin-related curvature on the orbital motion of
free electron related with topological features of the system and
thus is quite a general property. It can be used for detection of
inhomogeneous distribution of magnetic systems like
topological solitons and/or skyrmions recently observed in
AFMs.\cite{Raicev:PhysRevLett.106.227206} Adiabatic spin transport
is a possible tool for study of AFM 2D and 3D textures induced by
mechanical tilting through the flexomagnetic effect typical for
triangular AFM structures.\cite{Lukashev:PhysRevB.82.094417} A
 system of triangular Ising spins realized with trapped ions \cite{Kim:nature_2010}  can also be also used as a playground for quantum simulation of non-Abelian SU(2) Berry-phase effects.
\acknowledgements
I am grateful  to Yu. Mokrousov, F. Freimut and J. Sinova for fruitful discussions. I also acknowlegde help from P. Buhl and A. Malyshenko.
\appendix

\section{Hamiltonian diagonalization}

\label{app:diagonalization}

Hamiltonian (\ref{eq_hamiltonian}) mixes all six wavefunctions $\left\vert
u_{j}\right\rangle \left\vert \tau \right\rangle $,$j=1,2,3,\tau =\uparrow
,\downarrow $. However, symmetry considerations make it possible to simplify the
diagonalization procedure which we describe in this section.

First, we notice that if the quantization axis is taken in the local frame with $%
z_{j}^{\prime }$ axis parallel to $\mathbf{S}_{j}$, then, the $sd$-exchange-term
is diagonal in the spin space: $\mathbf{S}_{j}\hat{\boldsymbol{\sigma }}_{\tau
\tau ^{\prime }}=S\hat{\sigma}_{jz^{\prime }}$. As the vectors $\mathbf{S}_{j}$
(and correspondingly local quantization axes) could be generated from the
lab axis by rotations around the plane normal $\mathbf{n}$ it is convenient
to introduce new creation/annihilation operators
\begin{equation}
\left(\begin{array}{c}
\hat{b}_{j\uparrow }\\
\hat{b}_{j\downarrow }
\end{array}\right)
=\hat{U}_{j}^{\dag }%
\left(\begin{array}{c}
\hat{a}_{j\uparrow }\\
\hat{a}_{j\downarrow }
\end{array}\right),\qquad j=1,2,3,
\label{eq:transformation_first}
\end{equation}
where the unitary operators $\hat{U}_{j}=\cos \frac{\theta _{j}}{2}\hat{1}%
-i\sin \frac{\theta _{j}}{2}\mathbf{n}\hat{\boldsymbol{\sigma }}$ represent
rotation through the angles $\theta _{1}=0$, $\theta _{2}=2\pi /3$, $\theta
_{3}=4\pi /3$.

For further simplification we take the permutation symmetry of
the magnetic sublattices \cite{Gomonaj:Phase_Tr_1992a} into account and introduce the following
combinations of operators $\hat{b}_{j\tau }$
\begin{eqnarray}  \label{eq:transformation_second}
\hat{\xi}_{1\tau }&=&\frac{1}{\sqrt{6}}\left( 2\hat{b}_{1\tau }-\hat{b}%
_{2\tau }-\hat{b}_{3\tau }\right) ,\hat{\xi}_{2\tau }=\frac{1}{\sqrt{2}}%
\left( \hat{b}_{2\tau }-\hat{b}_{3\tau }\right),  \nonumber \\
\hat{\xi}_{3\tau }&=&\frac{1}{\sqrt{3}}\left( \hat{b}_{1\tau }+\hat{b}%
_{2\tau }+\hat{b}_{3\tau }\right)
\end{eqnarray}
that form irreducible
representations of permutation group P$_{3}$ (isomorphic to C$_{3}$ rotation
group which describes the exchange symmetry of the compound).

Operators $\left\{ \hat{b}_{j\tau },\hat{b}_{j\tau }^{\dag }\right\} $and $%
\left\{ \hat{\xi}_{j\tau },\hat{\xi}_{j\tau }^{\dag }\right\} $ satisfy the
same anticommutation relations as Fermi-operators $\left\{ \hat{a}_{j\tau },%
\hat{a}_{j\tau }^{\dag }\right\} $.

It is worth to mention that the operators $\left\{ \hat{\xi}_{1\tau },\hat{\xi}%
_{2\tau }\right\} $ belong to the same irreducible representation as the AFM
vectors $\left\{ \mathbf{L}_{1},\mathbf{L}_{2}\right\} $, and operator $%
\left\{ \hat{\xi}_{3\tau }\right\} $ has the same transformation properties
as magnetization vector $\mathbf{M}$ (see equation (\ref{eq:AFM_structure_mgnetization})).

As it was already mentioned, in the AFM ground state $\mathbf{M=0}$, $%
\mathbf{L}_{1}\bot \mathbf{L}_{2}\bot \mathbf{n,}\left\vert \mathbf{L}%
_{1}\right\vert =\left\vert \mathbf{L}_{2}\right\vert =S$.\cite%
{Gomonaj:Phase_Tr_1992a} We take the quantization axis for free spins parallel to $%
\mathbf{L}_{1}$.

Taking account of transformations (\ref{eq:transformation_first}), (\ref%
{eq:transformation_second}) the hamiltonian (\ref{eq_hamiltonian}) takes a form
\begin{eqnarray}
\hat{H}\left( \mathbf{r},t\right) &=&-J_{sd}S\sum_{j}\left( \hat{\xi}%
_{j\uparrow }^{\dag }\hat{\xi}_{j\uparrow }-\hat{\xi}_{j\downarrow }^{\dag }%
\hat{\xi}_{j\downarrow }\right)+ \gamma (\mathbf{k})\left[ 2\hat{\xi}_{3\tau
}^{\dag }\hat{\xi}_{3\tau }-\left( \hat{\xi}_{1\tau }^{\dag }\hat{\xi}%
_{1\tau }+\hat{\xi}_{2\tau }^{\dag }\hat{\xi}_{2\tau }\right) \right.  \nonumber
\\
&+&\left.3\left( \hat{\xi}_{1\downarrow }^{\dag }\hat{\xi}_{2\uparrow }+\hat{%
\xi}_{2\uparrow }^{\dag }\hat{\xi}_{1\downarrow }-\hat{\xi}_{2\downarrow
}^{\dag }\hat{\xi}_{1\uparrow }-\hat{\xi}_{1\uparrow }^{\dag }\hat{\xi}%
_{2\downarrow }\right) \right] ,
\end{eqnarray}
which now can be easily diagonalized.

\section{Berry connection, Berry curvature and dynamics equations}

\label{app:Berry_connection} To simplify the calculation of the Berry connection (%
\ref{eq:gauge_potential_def}) we note that i) time/space dependence of the
state vectors $|\Psi _{2a}(r_{\mu })\rangle $, $|\Psi _{2b}(r_{\mu })\rangle
$ stems from the rotation of the spin quantization axis; ii) rotation of the
local frame (unit vectors $\mathbf{e}_{k}(r_{\mu })$, $k=x,y,z$) can be
equivalently represented in terms of the rotation matrix $\Re $ (\ref%
{eq:Gibbs_vector}): $\mathbf{e}_{k}(r_{\mu })\equiv \Re (\boldsymbol{\varphi
})e_{k}^{0}$ or related unitary matrix $\hat{U}$ (\ref{eq:unitary
transformation}): $\mathbf{e}\boldsymbol{\hat{\sigma}=}\hat{U}\mathbf{e}^{0}%
\boldsymbol{\hat{\sigma}}\hat{U}^{\dag }$, where
$\mathbf{e}_{k}^{0}$ is taken at some reference point $r_{\mu
}^{0}$. The same matrix $\hat{U}$ defines transformation
(\ref{eq:transformaed state}) of state the vectors $|\Psi _{2}(r_{\mu
})\rangle $. An analogous procedure was proposed in
Ref.~\onlinecite{Tatara_PhysRevLett.78.3773}. So, the Berry connection
$\hat{\mathcal{A}}_{\mu }^{r}$ can be expressed as
\begin{equation}
\hat{\mathcal{A}}_{\mu }^{r}=i\left(
\begin{array}{cc}
\left\langle \Psi _{2a}^{0}\mid \hat{\Lambda}_{\mu }\Psi
_{2a}^{0}\right\rangle & \left\langle \Psi _{2a}^{0}\mid \hat{\Lambda}_{\mu
}\Psi _{2b}^{0}\right\rangle \\
\left\langle \Psi _{2b}^{0}\mid \hat{\Lambda}_{\mu }\Psi
_{2a}^{0}\right\rangle & \left\langle \Psi _{2b}^{0}\mid \hat{\Lambda}_{\mu
}\Psi _{2b}^{0}\right\rangle%
\end{array}%
\right) ,  \label{eq:gauge_potential_transf}
\end{equation}%
where we introduced the matrix $\hat{\Lambda}_{\mu }\equiv \hat{U}^{\dag
}\partial _{\mu }^{r}\hat{U}$. Direct calculations show that
\begin{equation}
\hat{\Lambda}_{\mu }=-\frac{i}{2}\Re ^{-1}(\boldsymbol{\varphi })\boldsymbol{%
\Omega }_{\mu }\hat{\boldsymbol{\sigma }}.  \label{eq:lambda}
\end{equation}%
and
\begin{equation}
\hat{\mathcal{A}}_{\mu }^{r}=\left( \frac{1}{2}\sin 2\psi \Re ^{-1}(%
\boldsymbol{\varphi })\boldsymbol{\Omega }_{\mu }+\frac{1}{4}\left( 1-\sin
2\psi \right) \mathbf{n}_{0}\times \Re ^{-1}(\boldsymbol{\varphi })\mathbf{%
\Omega }_{\mu }\times \mathbf{n}_{0}\right) \hat{\boldsymbol{\sigma }}.
\label{eq:Berry_connection_altimate}
\end{equation}%
From (\ref{eq:Berry_connection_altimate}) after some simple math one gets
expr. (\ref{eq:Berry_connection}).

It is obvious from relations (\ref{eq:gauge_potential_transf}), (\ref%
{eq:lambda}) that the singlet states with zero spin do not contribute to the Berry
connection.

The spin vector $\mathbf{s}(t,\mathbf{r)}=\left\langle w\right\vert \hat{%
\boldsymbol{\sigma }}\left\vert w\right\rangle $ is calculated in a similar
way with the help of matrix $\hat{\Sigma}\equiv \hat{U}^{\dag }\hat{%
\boldsymbol{\sigma }}\hat{U}=\Re (\boldsymbol{\varphi )e}_{k}^{0}\hat{\sigma
_{k}}$, explicit relation being
\begin{equation}
\mathbf{s(}t,\mathbf{r)=}\Re (\boldsymbol{\varphi )}\left[ \sin 2\psi
\mathbf{C}+\frac{1}{2}\left( 1-\sin 2\psi \right) \mathbf{n}_{0}\times
\left( \mathbf{C}\times \mathbf{n}_{0}\right) \right] .
\label{eq:spin_isospin_initial}
\end{equation}

Using the relation $\mathbf{n}=\Re (\boldsymbol{\varphi )}\mathbf{n}_{0}$
between the vectors in the local and reference frame one gets expression (\ref%
{eq:spin_isospin}) from (\ref{eq:spin_isospin_initial}).

Equation (\ref{eq:spin_isospin_initial}) is easily inverted in order to
express the isospin $\mathbf{C}$ through the real spin as follows:

\begin{equation}
 \mathbf{C=}\frac{2}{1+\sin 2\psi }\Re ^{-1}(\mathbf{\varphi )}\left[
\mathbf{s(}t,\mathbf{r)+}\frac{1-\sin 2\psi }{2\sin 2\psi }\left( \mathbf{ns}%
\right) \mathbf{n}\right]  \label{eq:isospin_spin}
\end{equation}

Substituting relation (\ref{eq:isospin_spin}) into the normalization condition $%
\left\vert \mathbf{C}\right\vert ^{2}=1$ one gets equation (\ref%
{eq:spin_normal relation}) for spheroid.

The Berry curvatures (\ref{eq:Berry_curvatures_2}) are calculated by
differentiation of (\ref{eq:Berry_connection_altimate}) with account of two
general relations:
\begin{equation}
\dot{\Re}(\boldsymbol{\varphi )=\Omega }\mathbf{\times }\Re (\boldsymbol{%
\varphi )~}\text{and}~\partial _{\nu }^{r}\boldsymbol{\Omega }_{\mu
}-\partial _{\mu }^{r}\mathbf{\Omega }_{\nu }=\boldsymbol{\Omega }_{\mu
}\times \boldsymbol{\Omega }_{\nu }.  \label{eq:useful_relations}
\end{equation}

Dynamic equation (\ref{eq:motions_1}) for the spin is obtained by
differentiation of the expression (\ref{eq:spin_isospin}) with the use of
relations (\ref{eq:useful_relations}), (\ref{eq_dynamics initial}) and (\ref{eq:isospin_spin}).

The complete equation for acceleration has the following form:
\begin{equation}
\dot{k}_{\mu }=-\partial _{\mu }^{r}\varepsilon_2 -\frac{1}{2}\mathbf{\dot{s}%
\cdot \mathbf{\Omega }}_{\mu }-\frac{1}{2}\dot{k}_{\nu }\partial _{\nu
}^{k}\ln \left( 1+\sin 2\psi \right) \left[ \mathbf{s\cdot \mathbf{\Omega }%
_{\mu }+}\frac{1}{\sin 2\psi }\left( \mathbf{ns}\right) \left( \mathbf{%
n\Omega }_{\mu }\right) \right] .  \label{eq:motion_acceleration_full}
\end{equation}


\begin{thebibliography}{51}%
\makeatletter
\providecommand \@ifxundefined [1]{%
 \@ifx{#1\undefined}
}%
\providecommand \@ifnum [1]{%
 \ifnum #1\expandafter \@firstoftwo
 \else \expandafter \@secondoftwo
 \fi
}%
\providecommand \@ifx [1]{%
 \ifx #1\expandafter \@firstoftwo
 \else \expandafter \@secondoftwo
 \fi
}%
\providecommand \natexlab [1]{#1}%
\providecommand \enquote  [1]{``#1''}%
\providecommand \bibnamefont  [1]{#1}%
\providecommand \bibfnamefont [1]{#1}%
\providecommand \citenamefont [1]{#1}%
\providecommand \href@noop [0]{\@secondoftwo}%
\providecommand \href [0]{\begingroup \@sanitize@url \@href}%
\providecommand \@href[1]{\@@startlink{#1}\@@href}%
\providecommand \@@href[1]{\endgroup#1\@@endlink}%
\providecommand \@sanitize@url [0]{\catcode `\\12\catcode
`\$12\catcode
  `\&12\catcode `\#12\catcode `\^12\catcode `\_12\catcode `\%12\relax}%
\providecommand \@@startlink[1]{}%
\providecommand \@@endlink[0]{}%
\providecommand \url  [0]{\begingroup\@sanitize@url \@url }%
\providecommand \@url [1]{\endgroup\@href {#1}{\urlprefix }}%
\providecommand \urlprefix  [0]{URL }%
\providecommand \Eprint [0]{\href }%
\providecommand \doibase [0]{http://dx.doi.org/}%
\providecommand \selectlanguage [0]{\@gobble}%
\providecommand \bibinfo  [0]{\@secondoftwo}%
\providecommand \bibfield  [0]{\@secondoftwo}%
\providecommand \translation [1]{[#1]}%
\providecommand \BibitemOpen [0]{}%
\providecommand \bibitemStop [0]{}%
\providecommand \bibitemNoStop [0]{.\EOS\space}%
\providecommand \EOS [0]{\spacefactor3000\relax}%
\providecommand \BibitemShut  [1]{\csname bibitem#1\endcsname}%
\let\auto@bib@innerbib\@empty
\bibitem [{\citenamefont {Tang}\ \emph {et~al.}(2009)\citenamefont {Tang},
  \citenamefont {Zhang}, \citenamefont {Sua}, \citenamefont {Jing},\ and\
  \citenamefont {Zhong}}]{Tang:2009}%
  \BibitemOpen
  \bibfield  {author} {\bibinfo {author} {\bibfnamefont {X.}~\bibnamefont
  {Tang}}, \bibinfo {author} {\bibfnamefont {H.-W.}\ \bibnamefont {Zhang}},
  \bibinfo {author} {\bibfnamefont {H.}~\bibnamefont {Sua}}, \bibinfo {author}
  {\bibfnamefont {Y.-L.}\ \bibnamefont {Jing}}, \ and\ \bibinfo {author}
  {\bibfnamefont {Z.-Y.}\ \bibnamefont {Zhong}},\ }\href {\doibase
  doi:10.1016/j.jmmm.2008.11.100} {\bibfield  {journal} {\bibinfo  {journal}
  {J. Magn. Mag. Mater.}\ }\textbf {\bibinfo {volume} {321}},\ \bibinfo {pages}
  {1851} (\bibinfo {year} {2009})}\BibitemShut {NoStop}%
\bibitem [{\citenamefont {Kampfrath}\ \emph {et~al.}(2011)\citenamefont
  {Kampfrath}, \citenamefont {Sell}, \citenamefont {Klatt}, \citenamefont
  {Pashkin}, \citenamefont {Mahrlein}, \citenamefont {Dekorsy}, \citenamefont
  {Wolf}, \citenamefont {Fiebig}, \citenamefont {Leitenstorfer},\ and\
  \citenamefont {Huber}}]{Kampfrath:2011}%
  \BibitemOpen
  \bibfield  {author} {\bibinfo {author} {\bibfnamefont {T.}~\bibnamefont
  {Kampfrath}}, \bibinfo {author} {\bibfnamefont {A.}~\bibnamefont {Sell}},
  \bibinfo {author} {\bibfnamefont {G.}~\bibnamefont {Klatt}}, \bibinfo
  {author} {\bibfnamefont {A.}~\bibnamefont {Pashkin}}, \bibinfo {author}
  {\bibfnamefont {S.}~\bibnamefont {Mahrlein}}, \bibinfo {author}
  {\bibfnamefont {T.}~\bibnamefont {Dekorsy}}, \bibinfo {author} {\bibfnamefont
  {M.}~\bibnamefont {Wolf}}, \bibinfo {author} {\bibfnamefont {M.}~\bibnamefont
  {Fiebig}}, \bibinfo {author} {\bibfnamefont {A.}~\bibnamefont
  {Leitenstorfer}}, \ and\ \bibinfo {author} {\bibfnamefont {R.}~\bibnamefont
  {Huber}},\ }\href {\doibase 10.1038/nphoton.2010.259} {\bibfield  {journal}
  {\bibinfo  {journal} {Nat Photon}\ }\textbf {\bibinfo {volume} {5}},\
  \bibinfo {pages} {31} (\bibinfo {year} {2011})}\BibitemShut {NoStop}%
\bibitem [{\citenamefont {Haney}\ \emph {et~al.}(2008)\citenamefont {Haney},
  \citenamefont {Duine}, \citenamefont {Nunez},\ and\ \citenamefont
  {MacDonald}}]{Haney:2008}%
  \BibitemOpen
  \bibfield  {author} {\bibinfo {author} {\bibfnamefont {P.}~\bibnamefont
  {Haney}}, \bibinfo {author} {\bibfnamefont {R.}~\bibnamefont {Duine}},
  \bibinfo {author} {\bibfnamefont {A.~S.}\ \bibnamefont {Nunez}}, \ and\
  \bibinfo {author} {\bibfnamefont {A.}~\bibnamefont {MacDonald}},\ }\href
  {\doibase 10.1016/j.jmmm.2007.12.020} {\bibfield  {journal} {\bibinfo
  {journal} {J. Magn. Magn. Mater.}\ }\textbf {\bibinfo {volume} {320}},\
  \bibinfo {pages} {1300} (\bibinfo {year} {2008})}\BibitemShut {NoStop}%
\bibitem [{\citenamefont {Gomonay}\ and\ \citenamefont
  {Loktev}(2008{\natexlab{a}})}]{Gomonay:2008_JMSJ}%
  \BibitemOpen
  \bibfield  {author} {\bibinfo {author} {\bibfnamefont {H.}~\bibnamefont
  {Gomonay}}\ and\ \bibinfo {author} {\bibfnamefont {V.}~\bibnamefont
  {Loktev}},\ }\href {\doibase http://dx.doi.org/10.3379/msjmag.32.535}
  {\bibfield  {journal} {\bibinfo  {journal} {J. Magn. Soc. .Japan}\ }\textbf
  {\bibinfo {volume} {32}},\ \bibinfo {pages} {535} (\bibinfo {year}
  {2008}{\natexlab{a}})}\BibitemShut {NoStop}%
\bibitem [{\citenamefont {Gomonay}\ and\ \citenamefont
  {Loktev}(2008{\natexlab{b}})}]{gomo:2008E}%
  \BibitemOpen
  \bibfield  {author} {\bibinfo {author} {\bibfnamefont {E.~V.}\ \bibnamefont
  {Gomonay}}\ and\ \bibinfo {author} {\bibfnamefont {V.~M.}\ \bibnamefont
  {Loktev}},\ }\href {\doibase 10.1063/1.2889408} {\bibfield  {journal}
  {\bibinfo  {journal} {Low Temperature Physics}\ }\textbf {\bibinfo {volume}
  {34}},\ \bibinfo {pages} {198} (\bibinfo {year}
  {2008}{\natexlab{b}})}\BibitemShut {NoStop}%
\bibitem [{\citenamefont {Cheng}\ \emph {et~al.}(2014)\citenamefont {Cheng},
  \citenamefont {Xiao}, \citenamefont {Niu},\ and\ \citenamefont
  {Brataas}}]{Cheng:PhysRevLett.113.057601}%
  \BibitemOpen
  \bibfield  {author} {\bibinfo {author} {\bibfnamefont {R.}~\bibnamefont
  {Cheng}}, \bibinfo {author} {\bibfnamefont {J.}~\bibnamefont {Xiao}},
  \bibinfo {author} {\bibfnamefont {Q.}~\bibnamefont {Niu}}, \ and\ \bibinfo
  {author} {\bibfnamefont {A.}~\bibnamefont {Brataas}},\ }\href {\doibase
  10.1103/PhysRevLett.113.057601} {\bibfield  {journal} {\bibinfo  {journal}
  {Phys. Rev. Lett.}\ }\textbf {\bibinfo {volume} {113}},\ \bibinfo {pages}
  {057601} (\bibinfo {year} {2014})}\BibitemShut {NoStop}%
\bibitem [{\citenamefont {Hals}\ \emph {et~al.}(2011)\citenamefont {Hals},
  \citenamefont {Tserkovnyak},\ and\ \citenamefont
  {Brataas}}]{Tserkovnyak:PhysRevLett.106.107206}%
  \BibitemOpen
  \bibfield  {author} {\bibinfo {author} {\bibfnamefont {K.~M.~D.}\
  \bibnamefont {Hals}}, \bibinfo {author} {\bibfnamefont {Y.}~\bibnamefont
  {Tserkovnyak}}, \ and\ \bibinfo {author} {\bibfnamefont {A.}~\bibnamefont
  {Brataas}},\ }\href {\doibase 10.1103/PhysRevLett.106.107206} {\bibfield
  {journal} {\bibinfo  {journal} {Phys. Rev. Lett.}\ }\textbf {\bibinfo
  {volume} {106}},\ \bibinfo {pages} {107206} (\bibinfo {year}
  {2011})}\BibitemShut {NoStop}%
\bibitem [{\citenamefont {Tveten}\ \emph {et~al.}(2013)\citenamefont {Tveten},
  \citenamefont {Qaiumzadeh}, \citenamefont {Tretiakov},\ and\ \citenamefont
  {Brataas}}]{Brataas:PhysRevLett.110.127208}%
  \BibitemOpen
  \bibfield  {author} {\bibinfo {author} {\bibfnamefont {E.~G.}\ \bibnamefont
  {Tveten}}, \bibinfo {author} {\bibfnamefont {A.}~\bibnamefont {Qaiumzadeh}},
  \bibinfo {author} {\bibfnamefont {O.~A.}\ \bibnamefont {Tretiakov}}, \ and\
  \bibinfo {author} {\bibfnamefont {A.}~\bibnamefont {Brataas}},\ }\href
  {\doibase 10.1103/PhysRevLett.110.127208} {\bibfield  {journal} {\bibinfo
  {journal} {Phys. Rev. Lett.}\ }\textbf {\bibinfo {volume} {110}},\ \bibinfo
  {pages} {127208} (\bibinfo {year} {2013})}\BibitemShut {NoStop}%
\bibitem [{\citenamefont {{\v{Z}}elezn{\'y}}\ \emph {et~al.}(2014)\citenamefont
  {{\v{Z}}elezn{\'y}}, \citenamefont {Gao}, \citenamefont {V\'yborn\'y},
  \citenamefont {Zemen}, \citenamefont {Ma{\v{s}}ek}, \citenamefont {Manchon},
  \citenamefont {Wunderlich}, \citenamefont {Sinova},\ and\ \citenamefont
  {Jungwirth}}]{Zelezny:PhysRevLett.113.157201}%
  \BibitemOpen
  \bibfield  {author} {\bibinfo {author} {\bibfnamefont {J.}~\bibnamefont
  {{\v{Z}}elezn{\'y}}}, \bibinfo {author} {\bibfnamefont {H.}~\bibnamefont
  {Gao}}, \bibinfo {author} {\bibfnamefont {K.}~\bibnamefont {V\'yborn\'y}},
  \bibinfo {author} {\bibfnamefont {J.}~\bibnamefont {Zemen}}, \bibinfo
  {author} {\bibfnamefont {J.}~\bibnamefont {Ma{\v{s}}ek}}, \bibinfo {author}
  {\bibfnamefont {A.}~\bibnamefont {Manchon}}, \bibinfo {author} {\bibfnamefont
  {J.}~\bibnamefont {Wunderlich}}, \bibinfo {author} {\bibfnamefont
  {J.}~\bibnamefont {Sinova}}, \ and\ \bibinfo {author} {\bibfnamefont
  {T.}~\bibnamefont {Jungwirth}},\ }\href {\doibase
  10.1103/PhysRevLett.113.157201} {\bibfield  {journal} {\bibinfo  {journal}
  {Phys. Rev. Lett.}\ }\textbf {\bibinfo {volume} {113}},\ \bibinfo {pages}
  {157201} (\bibinfo {year} {2014})}\BibitemShut {NoStop}%
\bibitem [{\citenamefont {Chen}\ \emph {et~al.}(2014)\citenamefont {Chen},
  \citenamefont {Niu},\ and\ \citenamefont
  {MacDonald}}]{MacDonald:PhysRevLett.112.017205}%
  \BibitemOpen
  \bibfield  {author} {\bibinfo {author} {\bibfnamefont {H.}~\bibnamefont
  {Chen}}, \bibinfo {author} {\bibfnamefont {Q.}~\bibnamefont {Niu}}, \ and\
  \bibinfo {author} {\bibfnamefont {A.}~\bibnamefont {MacDonald}},\ }\href
  {\doibase 10.1103/PhysRevLett.112.017205} {\bibfield  {journal} {\bibinfo
  {journal} {Phys. Rev. Lett.}\ }\textbf {\bibinfo {volume} {112}},\ \bibinfo
  {pages} {017205} (\bibinfo {year} {2014})}\BibitemShut {NoStop}%
\bibitem [{\citenamefont {{K{\"u}bler}}\ and\ \citenamefont
  {{Felser}}(2014)}]{Kubler:2014arXiv1410.5985K}%
  \BibitemOpen
  \bibfield  {author} {\bibinfo {author} {\bibfnamefont {J.}~\bibnamefont
  {{K{\"u}bler}}}\ and\ \bibinfo {author} {\bibfnamefont {C.}~\bibnamefont
  {{Felser}}},\ }\href@noop {} {\bibfield  {journal} {\bibinfo  {journal}
  {ArXiv e-prints}\ } (\bibinfo {year} {2014})},\ \Eprint
  {http://arxiv.org/abs/1410.5985} {arXiv:1410.5985}
  \BibitemShut {NoStop}%
\bibitem [{\citenamefont {Bruno}\ \emph {et~al.}(2004)\citenamefont {Bruno},
  \citenamefont {Dugaev},\ and\ \citenamefont
  {Taillefumier}}]{Bruno:PhysRevLett.93.096806}%
  \BibitemOpen
  \bibfield  {author} {\bibinfo {author} {\bibfnamefont {P.}~\bibnamefont
  {Bruno}}, \bibinfo {author} {\bibfnamefont {V.~K.}\ \bibnamefont {Dugaev}}, \
  and\ \bibinfo {author} {\bibfnamefont {M.}~\bibnamefont {Taillefumier}},\
  }\href {\doibase 10.1103/PhysRevLett.93.096806} {\bibfield  {journal}
  {\bibinfo  {journal} {Phys. Rev. Lett.}\ }\textbf {\bibinfo {volume} {93}},\
  \bibinfo {pages} {096806} (\bibinfo {year} {2004})}\BibitemShut {NoStop}%
\bibitem [{\citenamefont {Nagaosa}\ \emph {et~al.}(2010)\citenamefont
  {Nagaosa}, \citenamefont {Sinova}, \citenamefont {Onoda}, \citenamefont
  {MacDonald},\ and\ \citenamefont {Ong}}]{MacDonald:RevModPhys.82.1539}%
  \BibitemOpen
  \bibfield  {author} {\bibinfo {author} {\bibfnamefont {N.}~\bibnamefont
  {Nagaosa}}, \bibinfo {author} {\bibfnamefont {J.}~\bibnamefont {Sinova}},
  \bibinfo {author} {\bibfnamefont {S.}~\bibnamefont {Onoda}}, \bibinfo
  {author} {\bibfnamefont {A.~H.}\ \bibnamefont {MacDonald}}, \ and\ \bibinfo
  {author} {\bibfnamefont {N.~P.}\ \bibnamefont {Ong}},\ }\href {\doibase
  10.1103/RevModPhys.82.1539} {\bibfield  {journal} {\bibinfo  {journal} {Rev.
  Mod. Phys.}\ }\textbf {\bibinfo {volume} {82}},\ \bibinfo {pages} {1539}
  (\bibinfo {year} {2010})}\BibitemShut {NoStop}%
\bibitem [{\citenamefont {Takatsu}\ \emph {et~al.}(2010)\citenamefont
  {Takatsu}, \citenamefont {Yonezawa}, \citenamefont {Fujimoto},\ and\
  \citenamefont {Maeno}}]{Takatsu:PhysRevLett.105.137201}%
  \BibitemOpen
  \bibfield  {author} {\bibinfo {author} {\bibfnamefont {H.}~\bibnamefont
  {Takatsu}}, \bibinfo {author} {\bibfnamefont {S.}~\bibnamefont {Yonezawa}},
  \bibinfo {author} {\bibfnamefont {S.}~\bibnamefont {Fujimoto}}, \ and\
  \bibinfo {author} {\bibfnamefont {Y.}~\bibnamefont {Maeno}},\ }\href
  {\doibase 10.1103/PhysRevLett.105.137201} {\bibfield  {journal} {\bibinfo
  {journal} {Phys. Rev. Lett.}\ }\textbf {\bibinfo {volume} {105}},\ \bibinfo
  {pages} {137201} (\bibinfo {year} {2010})}\BibitemShut {NoStop}%
\bibitem [{\citenamefont {S{\"{u}}rgers}\ \emph {et~al.}(2014)\citenamefont
  {S{\"{u}}rgers}, \citenamefont {Fischer}, \citenamefont {Winkel},\ and\
  \citenamefont {v.~L\"{o}hneysen}}]{Christoph_Surgers:2014}%
  \BibitemOpen
  \bibfield  {author} {\bibinfo {author} {\bibfnamefont {C.}~\bibnamefont
  {S{\"{u}}rgers}}, \bibinfo {author} {\bibfnamefont {G.}~\bibnamefont
  {Fischer}}, \bibinfo {author} {\bibfnamefont {P.}~\bibnamefont {Winkel}}, \
  and\ \bibinfo {author} {\bibfnamefont {H.}~\bibnamefont {v.~L\"{o}hneysen}},\
  }\href {\doibase 10.1038/ncomms4400} {\bibfield  {journal} {\bibinfo
  {journal} {Nature Comm.}\ }\textbf {\bibinfo {volume} {5}},\ \bibinfo {pages}
  {3400} (\bibinfo {year} {2014})}\BibitemShut {NoStop}%
\bibitem [{\citenamefont {Tserkovnyak}\ and\ \citenamefont
  {Wong}(2009)}]{Tserkovnyak:PhysRevB.79.014402}%
  \BibitemOpen
  \bibfield  {author} {\bibinfo {author} {\bibfnamefont {Y.}~\bibnamefont
  {Tserkovnyak}}\ and\ \bibinfo {author} {\bibfnamefont {C.~H.}\ \bibnamefont
  {Wong}},\ }\href {\doibase 10.1103/PhysRevB.79.014402} {\bibfield  {journal}
  {\bibinfo  {journal} {Phys. Rev. B}\ }\textbf {\bibinfo {volume} {79}},\
  \bibinfo {pages} {014402} (\bibinfo {year} {2009})}\BibitemShut {NoStop}%
\bibitem [{\citenamefont {Cheng}\ and\ \citenamefont
  {Niu}(2012)}]{Cheng:PhysRevB.86.245118}%
  \BibitemOpen
  \bibfield  {author} {\bibinfo {author} {\bibfnamefont {R.}~\bibnamefont
  {Cheng}}\ and\ \bibinfo {author} {\bibfnamefont {Q.}~\bibnamefont {Niu}},\
  }\href {\doibase 10.1103/PhysRevB.86.245118} {\bibfield  {journal} {\bibinfo
  {journal} {Phys. Rev. B}\ }\textbf {\bibinfo {volume} {86}},\ \bibinfo
  {pages} {245118} (\bibinfo {year} {2012})}\BibitemShut {NoStop}%
\bibitem [{\citenamefont {Barnes}\ and\ \citenamefont
  {Maekawa}(2007)}]{Barnes:PhysRevLett.98.246601}%
  \BibitemOpen
  \bibfield  {author} {\bibinfo {author} {\bibfnamefont {S.~E.}\ \bibnamefont
  {Barnes}}\ and\ \bibinfo {author} {\bibfnamefont {S.}~\bibnamefont
  {Maekawa}},\ }\href {\doibase 10.1103/PhysRevLett.98.246601} {\bibfield
  {journal} {\bibinfo  {journal} {Phys. Rev. Lett.}\ }\textbf {\bibinfo
  {volume} {98}},\ \bibinfo {pages} {246601} (\bibinfo {year}
  {2007})}\BibitemShut {NoStop}%
\bibitem [{\citenamefont {Freimuth}\ \emph {et~al.}(2013)\citenamefont
  {Freimuth}, \citenamefont {Bamler}, \citenamefont {Mokrousov},\ and\
  \citenamefont {Rosch}}]{Mokrousov:PhysRevB.88.214409}%
  \BibitemOpen
  \bibfield  {author} {\bibinfo {author} {\bibfnamefont {F.}~\bibnamefont
  {Freimuth}}, \bibinfo {author} {\bibfnamefont {R.}~\bibnamefont {Bamler}},
  \bibinfo {author} {\bibfnamefont {Y.}~\bibnamefont {Mokrousov}}, \ and\
  \bibinfo {author} {\bibfnamefont {A.}~\bibnamefont {Rosch}},\ }\href
  {\doibase 10.1103/PhysRevB.88.214409} {\bibfield  {journal} {\bibinfo
  {journal} {Phys. Rev. B}\ }\textbf {\bibinfo {volume} {88}},\ \bibinfo
  {pages} {214409} (\bibinfo {year} {2013})}\BibitemShut {NoStop}%
\bibitem [{\citenamefont {Bertaut}\ \emph {et~al.}(1968)\citenamefont
  {Bertaut}, \citenamefont {Fruchart}, \citenamefont {Bouchaud},\ and\
  \citenamefont {Fruchart}}]{Bertaut1968251}%
  \BibitemOpen
  \bibfield  {author} {\bibinfo {author} {\bibfnamefont {E.}~\bibnamefont
  {Bertaut}}, \bibinfo {author} {\bibfnamefont {D.}~\bibnamefont {Fruchart}},
  \bibinfo {author} {\bibfnamefont {J.}~\bibnamefont {Bouchaud}}, \ and\
  \bibinfo {author} {\bibfnamefont {R.}~\bibnamefont {Fruchart}},\ }\href
  {\doibase http://dx.doi.org/10.1016/0038-1098(68)90098-7} {\bibfield
  {journal} {\bibinfo  {journal} {Solid State Comm.}\ }\textbf
  {\bibinfo {volume} {6}},\ \bibinfo {pages} {251 } (\bibinfo {year}
  {1968})}\BibitemShut {NoStop}%
\bibitem [{\citenamefont {{Jardin, J.P.}}\ and\ \citenamefont {{Labb\'{e},
  J.}}(1975)}]{Jardin:1975}%
  \BibitemOpen
  \bibfield  {author} {\bibinfo {author} {\bibnamefont {{J.P.}~{Jardin}}}\ and\
  \bibinfo {author} {\bibnamefont {{J.}~{Labb\'{e}}}},\ }\href {\doibase
  10.1051/jphys:0197500360120131700} {\bibfield  {journal} {\bibinfo  {journal}
  {J. Phys. France}\ }\textbf {\bibinfo {volume} {36}},\ \bibinfo {pages}
  {1317} (\bibinfo {year} {1975})}\BibitemShut {NoStop}%
\bibitem [{\citenamefont {Ali}\ \emph {et~al.}(2014)\citenamefont {Ali},
  \citenamefont {Shafiq}, \citenamefont {Asadabadi}, \citenamefont {Aliabad},
  \citenamefont {Khan},\ and\ \citenamefont {Ahmad}}]{Ali2014141}%
  \BibitemOpen
  \bibfield  {author} {\bibinfo {author} {\bibfnamefont {Z.}~\bibnamefont
  {Ali}}, \bibinfo {author} {\bibfnamefont {M.}~\bibnamefont {Shafiq}},
  \bibinfo {author} {\bibfnamefont {S.~J.}\ \bibnamefont {Asadabadi}}, \bibinfo
  {author} {\bibfnamefont {H.~R.}\ \bibnamefont {Aliabad}}, \bibinfo {author}
  {\bibfnamefont {I.}~\bibnamefont {Khan}}, \ and\ \bibinfo {author}
  {\bibfnamefont {I.}~\bibnamefont {Ahmad}},\ }\href {\doibase
  http://dx.doi.org/10.1016/j.commatsci.2013.07.040} {\bibfield  {journal}
  {\bibinfo  {journal} {Computational Materials Science}\ }\textbf {\bibinfo
  {volume} {81}},\ \bibinfo {pages} {141 } (\bibinfo {year}
  {2014})}\BibitemShut {NoStop}%
\bibitem [{\citenamefont {Motizuki}\ and\ \citenamefont
  {Nagai}(1988)}]{Motizuki:0022-3719-21-30-011}%
  \BibitemOpen
  \bibfield  {author} {\bibinfo {author} {\bibfnamefont {K.}~\bibnamefont
  {Motizuki}}\ and\ \bibinfo {author} {\bibfnamefont {H.}~\bibnamefont
  {Nagai}},\ }\href {http://stacks.iop.org/0022-3719/21/i=30/a=011} {\bibfield
  {journal} {\bibinfo  {journal} {J. Phys. C}\ }\textbf {\bibinfo {volume}
  {21}},\ \bibinfo {pages} {5251} (\bibinfo {year} {1988})}\BibitemShut
  {NoStop}%
\bibitem [{\citenamefont {Gomonaj}\ and\ \citenamefont
  {L'vov}(1992{\natexlab{a}})}]{Gomonaj:Phase_Tr_1992a}%
  \BibitemOpen
  \bibfield  {author} {\bibinfo {author} {\bibfnamefont {E.~V.}\ \bibnamefont
  {Gomonaj}}\ and\ \bibinfo {author} {\bibfnamefont {V.~A.}\ \bibnamefont
  {L'vov}},\ }\href {\doibase 10.1080/01411599208203457} {\bibfield  {journal}
  {\bibinfo  {journal} {Phase Transitions}\ }\textbf {\bibinfo {volume} {38}},\
  \bibinfo {pages} {15} (\bibinfo {year} {1992}{\natexlab{a}})} \BibitemShut {NoStop}%
\bibitem [{\citenamefont {Gomonaj}\ and\ \citenamefont
  {L'vov}(1992{\natexlab{b}})}]{Gomonaj:Phase_Tr_1992}%
  \BibitemOpen
  \bibfield  {author} {\bibinfo {author} {\bibfnamefont {E.~V.}\ \bibnamefont
  {Gomonaj}}\ and\ \bibinfo {author} {\bibfnamefont {V.~A.}\ \bibnamefont
  {L'vov}},\ }\href {\doibase 10.1080/01411599208207749} {\bibfield  {journal}
  {\bibinfo  {journal} {Phase Transitions}\ }\textbf {\bibinfo {volume} {40}},\
  \bibinfo {pages} {225} (\bibinfo {year} {1992}{\natexlab{b}})} \BibitemShut {NoStop}%
\bibitem [{\citenamefont {Lukashev}\ and\ \citenamefont
  {Sabirianov}(2010)}]{Lukashev:PhysRevB.82.094417}%
  \BibitemOpen
  \bibfield  {author} {\bibinfo {author} {\bibfnamefont {P.}~\bibnamefont
  {Lukashev}}\ and\ \bibinfo {author} {\bibfnamefont {R.~F.}\ \bibnamefont
  {Sabirianov}},\ }\href {\doibase 10.1103/PhysRevB.82.094417} {\bibfield
  {journal} {\bibinfo  {journal} {Phys. Rev. B}\ }\textbf {\bibinfo {volume}
  {82}},\ \bibinfo {pages} {094417} (\bibinfo {year} {2010})}\BibitemShut
  {NoStop}%
\bibitem [{\citenamefont {Chi}\ \emph {et~al.}(2001)\citenamefont {Chi},
  \citenamefont {Kim},\ and\ \citenamefont {Hur}}]{Chi2001307}%
  \BibitemOpen
  \bibfield  {author} {\bibinfo {author} {\bibfnamefont {E.}~\bibnamefont
  {Chi}}, \bibinfo {author} {\bibfnamefont {W.}~\bibnamefont {Kim}}, \ and\
  \bibinfo {author} {\bibfnamefont {N.}~\bibnamefont {Hur}},\ }\href {\doibase
  http://dx.doi.org/10.1016/S0038-1098(01)00395-7} {\bibfield  {journal}
  {\bibinfo  {journal} {Solid State Comm.}\ }\textbf {\bibinfo {volume}
  {120}},\ \bibinfo {pages} {307 } (\bibinfo {year} {2001})}\BibitemShut
  {NoStop}%
\bibitem [{\citenamefont {Takenaka}\ \emph {et~al.}(2011)\citenamefont
  {Takenaka}, \citenamefont {Ozawa}, \citenamefont {Shibayama}, \citenamefont
  {Kaneko}, \citenamefont {Oe},\ and\ \citenamefont
  {Urano}}]{Takenaka:apl/98/2/10.1063/1.3541449}%
  \BibitemOpen
  \bibfield  {author} {\bibinfo {author} {\bibfnamefont {K.}~\bibnamefont
  {Takenaka}}, \bibinfo {author} {\bibfnamefont {A.}~\bibnamefont {Ozawa}},
  \bibinfo {author} {\bibfnamefont {T.}~\bibnamefont {Shibayama}}, \bibinfo
  {author} {\bibfnamefont {N.}~\bibnamefont {Kaneko}}, \bibinfo {author}
  {\bibfnamefont {T.}~\bibnamefont {Oe}}, \ and\ \bibinfo {author}
  {\bibfnamefont {C.}~\bibnamefont {Urano}},\ }\href {\doibase
  http://dx.doi.org/10.1063/1.3541449} {\bibfield  {journal} {\bibinfo
  {journal} {Appl. Phys. Lett.}\ }\textbf {\bibinfo {volume} {98}},\
  \bibinfo {eid} {022103} (\bibinfo {year} {2011})}\BibitemShut {NoStop}%
\bibitem [{Note1()}]{Note1}%
  \BibitemOpen
  \bibinfo {note} {Stricktly speaking, $\Gamma ^{4g}$ structure realized in a
  certain temperature range is consistent with weak ferromagnetic structure in
  which vectors $\protect \mathbf {S}_{j}$ slightly deviate from the plane. We,
  however, neglect small noncompensated magnetization for the sake of
  simplicity.}\BibitemShut {Stop}%
\bibitem [{\citenamefont {Andreev}\ and\ \citenamefont
  {Marchenko}(1980)}]{Andreev:1980}%
  \BibitemOpen
  \bibfield  {author} {\bibinfo {author} {\bibfnamefont {A.~F.}\ \bibnamefont
  {Andreev}}\ and\ \bibinfo {author} {\bibfnamefont {V.~I.}\ \bibnamefont
  {Marchenko}},\ }\href {\doibase 10.1070/PU1980v023n01ABEH004859} {\bibfield
  {journal} {\bibinfo  {journal} {Physics-Uspekhi}\ }\textbf {\bibinfo {volume}
  {23}},\ \bibinfo {pages} {21} (\bibinfo {year} {1980})}\BibitemShut {NoStop}%
\bibitem [{\citenamefont {Gomonay}\ \emph {et~al.}(2012)\citenamefont
  {Gomonay}, \citenamefont {Kunitsyn},\ and\ \citenamefont
  {Loktev}}]{Gomonay:PhysRevB.85.134446}%
  \BibitemOpen
  \bibfield  {author} {\bibinfo {author} {\bibfnamefont {H.~V.}\ \bibnamefont
  {Gomonay}}, \bibinfo {author} {\bibfnamefont {R.~V.}\ \bibnamefont
  {Kunitsyn}}, \ and\ \bibinfo {author} {\bibfnamefont {V.~M.}\ \bibnamefont
  {Loktev}},\ }\href {\doibase 10.1103/PhysRevB.85.134446} {\bibfield
  {journal} {\bibinfo  {journal} {Phys. Rev. B}\ }\textbf {\bibinfo {volume}
  {85}},\ \bibinfo {pages} {134446} (\bibinfo {year} {2012})} \BibitemShut {NoStop}%
\bibitem [{\citenamefont {Chu}\ \emph {et~al.}(2012)\citenamefont {Chu},
  \citenamefont {Wang}, \citenamefont {Yan}, \citenamefont {Na}, \citenamefont
  {Ding}, \citenamefont {Sun},\ and\ \citenamefont {Wen}}]{Chu2012173}%
  \BibitemOpen
  \bibfield  {author} {\bibinfo {author} {\bibfnamefont {L.}~\bibnamefont
  {Chu}}, \bibinfo {author} {\bibfnamefont {C.}~\bibnamefont {Wang}}, \bibinfo
  {author} {\bibfnamefont {J.}~\bibnamefont {Yan}}, \bibinfo {author}
  {\bibfnamefont {Y.}~\bibnamefont {Na}}, \bibinfo {author} {\bibfnamefont
  {L.}~\bibnamefont {Ding}}, \bibinfo {author} {\bibfnamefont {Y.}~\bibnamefont
  {Sun}}, \ and\ \bibinfo {author} {\bibfnamefont {Y.}~\bibnamefont {Wen}},\
  }\href {\doibase http://dx.doi.org/10.1016/j.scriptamat.2012.04.008}
  {\bibfield  {journal} {\bibinfo  {journal} {Scripta Mater.}\ }\textbf
  {\bibinfo {volume} {67}},\ \bibinfo {pages} {173 } (\bibinfo {year}
  {2012})}\BibitemShut {NoStop}%
\bibitem [{\citenamefont {Sundaram}\ and\ \citenamefont
  {Niu}(1999)}]{Sundaram:PhysRevB.59.14915}%
  \BibitemOpen
  \bibfield  {author} {\bibinfo {author} {\bibfnamefont {G.}~\bibnamefont
  {Sundaram}}\ and\ \bibinfo {author} {\bibfnamefont {Q.}~\bibnamefont {Niu}},\
  }\href {\doibase 10.1103/PhysRevB.59.14915} {\bibfield  {journal} {\bibinfo
  {journal} {Phys. Rev. B}\ }\textbf {\bibinfo {volume} {59}},\ \bibinfo
  {pages} {14915} (\bibinfo {year} {1999})}\BibitemShut {NoStop}%
\bibitem [{\citenamefont {Xiao}\ \emph {et~al.}(2010)\citenamefont {Xiao},
  \citenamefont {Chang},\ and\ \citenamefont {Niu}}]{Xiao:RevModPhys.82.1959}%
  \BibitemOpen
  \bibfield  {author} {\bibinfo {author} {\bibfnamefont {D.}~\bibnamefont
  {Xiao}}, \bibinfo {author} {\bibfnamefont {M.-C.}\ \bibnamefont {Chang}}, \
  and\ \bibinfo {author} {\bibfnamefont {Q.}~\bibnamefont {Niu}},\ }\href
  {\doibase 10.1103/RevModPhys.82.1959} {\bibfield  {journal} {\bibinfo
  {journal} {Rev. Mod. Phys.}\ }\textbf {\bibinfo {volume} {82}},\ \bibinfo
  {pages} {1959} (\bibinfo {year} {2010})}\BibitemShut {NoStop}%
\bibitem [{\citenamefont {Culcer}\ \emph {et~al.}(2005)\citenamefont {Culcer},
  \citenamefont {Yao},\ and\ \citenamefont {Niu}}]{Culcer:PhysRevB.72.085110}%
  \BibitemOpen
  \bibfield  {author} {\bibinfo {author} {\bibfnamefont {D.}~\bibnamefont
  {Culcer}}, \bibinfo {author} {\bibfnamefont {Y.}~\bibnamefont {Yao}}, \ and\
  \bibinfo {author} {\bibfnamefont {Q.}~\bibnamefont {Niu}},\ }\href {\doibase
  10.1103/PhysRevB.72.085110} {\bibfield  {journal} {\bibinfo  {journal} {Phys.
  Rev. B}\ }\textbf {\bibinfo {volume} {72}},\ \bibinfo {pages} {085110}
  (\bibinfo {year} {2005})}\BibitemShut {NoStop}%
\bibitem [{Note2()}]{Note2}%
  \BibitemOpen
  \bibinfo {note} {~Note, that we discuss only rotations in spin state, the
  lattice itself is supposed to be unchanged. However, the theory can be
  generalized to include space rotations of lattice, as will be discussed
  below. See also Ref.\onlinecite{Sundaram:PhysRevB.59.14915} for description of gauge
  fields in the deformed crystals.}\BibitemShut {Stop}%
\bibitem [{\citenamefont {Tserkovnyak}\ and\ \citenamefont
  {Mecklenburg}(2008)}]{Tserkovnyak:PhysRevB.77.134407}%
  \BibitemOpen
  \bibfield  {author} {\bibinfo {author} {\bibfnamefont {Y.}~\bibnamefont
  {Tserkovnyak}}\ and\ \bibinfo {author} {\bibfnamefont {M.}~\bibnamefont
  {Mecklenburg}},\ }\href {\doibase 10.1103/PhysRevB.77.134407} {\bibfield
  {journal} {\bibinfo  {journal} {Phys. Rev. B}\ }\textbf {\bibinfo {volume}
  {77}},\ \bibinfo {pages} {134407} (\bibinfo {year} {2008})}\BibitemShut
  {NoStop}%
\bibitem [{\citenamefont {Nagaosa}\ and\ \citenamefont
  {Tokura}(2013)}]{Nagaosa:nnano.2013.243}%
  \BibitemOpen
  \bibfield  {author} {\bibinfo {author} {\bibfnamefont {N.}~\bibnamefont
  {Nagaosa}}\ and\ \bibinfo {author} {\bibfnamefont {Y.}~\bibnamefont
  {Tokura}},\ }\href {\doibase 10.1038/nnano.2013.243} {\bibfield  {journal}
  {\bibinfo  {journal} {Nat Nano}\ }\textbf {\bibinfo {volume} {8}},\ \bibinfo
  {pages} {899} (\bibinfo {year} {2013})}\BibitemShut {NoStop}%
  \bibitem [{\citenamefont {Volovik}(1987)}]{Volovik:0022-3719-20-7-003}%
    \BibitemOpen
    \bibfield  {author} {\bibinfo {author} {\bibfnamefont {G.~E.}\ \bibnamefont
    {Volovik}},\ }\href {http://stacks.iop.org/0022-3719/20/i=7/a=003} {\bibfield
     {journal} {\bibinfo  {journal} {J. Phys. C}\ }\textbf {\bibinfo {volume}
    {20}},\ \bibinfo {pages} {L83} (\bibinfo {year} {1987})}\BibitemShut
    {NoStop}%
  \bibitem [{\citenamefont {A.M.Kosevich}\ \emph {et~al.}(1983)\citenamefont
    {A.M.Kosevich}, \citenamefont {B.A.Ivanov},\ and\ \citenamefont
    {A.S.Kovalev}}]{Ivanov:1983E}%
    \BibitemOpen
    \bibfield  {author} {\bibinfo {author} {\bibnamefont {A.M.Kosevich}},
    \bibinfo {author} {\bibnamefont {B.A.Ivanov}}, \ and\ \bibinfo {author}
    {\bibnamefont {A.S.Kovalev}},\ }\href@noop {} {\emph {\bibinfo {title}
    {Nonlinear magnetization waves. Dynamical and topological solitons}}}\
    (\bibinfo  {publisher} {Naukova Dumka},\ \bibinfo {address} {Kiev},\ \bibinfo
    {year} {1983})\ \bibinfo {note} {189 p.(in Russian)}\BibitemShut {NoStop}%
    \bibitem [{\citenamefont {Rajaraman}()}]{Rajaraman}%
      \BibitemOpen
      \bibfield  {author} {\bibinfo {author} {\bibfnamefont {R.}~\bibnamefont
      {Rajaraman}},\ }\href@noop {} {\emph {\bibinfo {title} {Solitons And
      Instantons: An Introduction To Solitons And Instantons In Quantum Field
      Theory}}}\ (\bibinfo  {publisher} {North-Holland, Amsterdam})\BibitemShut
      {NoStop}%
\bibitem [{\citenamefont {Bazaliy}\ \emph {et~al.}(1998)\citenamefont
  {Bazaliy}, \citenamefont {Jones},\ and\ \citenamefont
  {Zhang}}]{Bazaliy:PhysRevB.57.R3213}%
  \BibitemOpen
  \bibfield  {author} {\bibinfo {author} {\bibfnamefont {Y.~B.}\ \bibnamefont
  {Bazaliy}}, \bibinfo {author} {\bibfnamefont {B.~A.}\ \bibnamefont {Jones}},
  \ and\ \bibinfo {author} {\bibfnamefont {S.-C.}\ \bibnamefont {Zhang}},\
  }\href {\doibase 10.1103/PhysRevB.57.R3213} {\bibfield  {journal} {\bibinfo
  {journal} {Phys. Rev. B}\ }\textbf {\bibinfo {volume} {57}},\ \bibinfo
  {pages} {R3213} (\bibinfo {year} {1998})}\BibitemShut {NoStop}%
\bibitem [{Note3()}]{Note3}%
  \BibitemOpen
  \bibinfo {note} {Strictly speaking, an effective charge should be introduced
  as a vector quantity, so, for the sake of simplicity we introduce combination
  of charge and fields. However, Faraday's relation for fields {$\varepsilon
  _{\xi \mu \nu }\partial _{\mu }E_{\nu }+\partial _{t}B_{\xi }=0$} is
  satisfied.}\BibitemShut {Stop}%
\bibitem [{\citenamefont {Shindou}\ and\ \citenamefont
  {Nagaosa}(2001)}]{Shindou:PhysRevLett.87.116801}%
  \BibitemOpen
  \bibfield  {author} {\bibinfo {author} {\bibfnamefont {R.}~\bibnamefont
  {Shindou}}\ and\ \bibinfo {author} {\bibfnamefont {N.}~\bibnamefont
  {Nagaosa}},\ }\href {\doibase 10.1103/PhysRevLett.87.116801} {\bibfield
  {journal} {\bibinfo  {journal} {Phys. Rev. Lett.}\ }\textbf {\bibinfo
  {volume} {87}},\ \bibinfo {pages} {116801} (\bibinfo {year}
  {2001})}\BibitemShut {NoStop}%
\bibitem [{Note4()}]{Note4}%
  \BibitemOpen
  \bibinfo {note} {Formally, it is combination $q\protect \mathbf {B}$ that has
  pseudo-vector character, due to the dependence on pseudovector $\protect
  \mathbf {s}$ (see (\ref {eq:magnetic_field})). So, in our case time-inversion
  symmetry is broken rather by the effective charge than by the effective
  field.}\BibitemShut {Stop}%
\bibitem [{\citenamefont {{Mart{\'{\i}}}}\ \emph {et~al.}(2012)\citenamefont
  {{Mart{\'{\i}}}}, \citenamefont {{Park}}, \citenamefont {{Wunderlich}},
  \citenamefont {{Reichlov{\'a}}}, \citenamefont {{Kurosaki}}, \citenamefont
  {{Yamada}}, \citenamefont {{Yamamoto}}, \citenamefont {{Nishide}},
  \citenamefont {{Hayakawa}}, \citenamefont {{Takahashi}},\ and\ \citenamefont
  {{Jungwirth}}}]{Jungwirth:2012PhRvL.108a7201M}%
  \BibitemOpen
  \bibfield  {author} {\bibinfo {author} {\bibfnamefont {X.}~\bibnamefont
  {{Mart{\'{\i}}}}}, \bibinfo {author} {\bibfnamefont {B.~G.}\ \bibnamefont
  {{Park}}}, \bibinfo {author} {\bibfnamefont {J.}~\bibnamefont
  {{Wunderlich}}}, \bibinfo {author} {\bibfnamefont {H.}~\bibnamefont
  {{Reichlov{\'a}}}}, \bibinfo {author} {\bibfnamefont {Y.}~\bibnamefont
  {{Kurosaki}}}, \bibinfo {author} {\bibfnamefont {M.}~\bibnamefont
  {{Yamada}}}, \bibinfo {author} {\bibfnamefont {H.}~\bibnamefont
  {{Yamamoto}}}, \bibinfo {author} {\bibfnamefont {A.}~\bibnamefont
  {{Nishide}}}, \bibinfo {author} {\bibfnamefont {J.}~\bibnamefont
  {{Hayakawa}}}, \bibinfo {author} {\bibfnamefont {H.}~\bibnamefont
  {{Takahashi}}}, \ and\ \bibinfo {author} {\bibfnamefont {T.}~\bibnamefont
  {{Jungwirth}}},\ }\href {\doibase 10.1103/PhysRevLett.108.017201} {\bibfield
  {journal} {\bibinfo  {journal} {Phys. Rev. Lett.}\ }\textbf {\bibinfo
  {volume} {108}},\ \bibinfo {eid} {017201} (\bibinfo {year} {2012})}\BibitemShut {NoStop}%
\bibitem [{Note5()}]{Note5}%
  \BibitemOpen
  \bibinfo {note} {The proposed model distribution of AFM order parameter has a
  peculiarity at $r\rightarrow 0$ which can be avoided by placing a defect or a
  hole in the center.}\BibitemShut {Stop}%
\bibitem [{Note6()}]{Note6}%
  \BibitemOpen
  \bibinfo {note} {Spin rotation which arises during the passage through the
  domain wall, can reduce the deflection. However, due to $1/r$ dependency of
  the Lorentz force the effect can be still pronounced close to the center of
  the region.}\BibitemShut {Stop}%
\bibitem [{\citenamefont {Lin}\ \emph {et~al.}(2011)\citenamefont {Lin},
  \citenamefont {Wang}, \citenamefont {Tong}, \citenamefont {Lin},
  \citenamefont {Lu}, \citenamefont {Zhu}, \citenamefont {Yang}, \citenamefont
  {Song}, \citenamefont {Dai},\ and\ \citenamefont {Sun}}]{Lin2011452}%
  \BibitemOpen
  \bibfield  {author} {\bibinfo {author} {\bibfnamefont {J.}~\bibnamefont
  {Lin}}, \bibinfo {author} {\bibfnamefont {B.}~\bibnamefont {Wang}}, \bibinfo
  {author} {\bibfnamefont {P.}~\bibnamefont {Tong}}, \bibinfo {author}
  {\bibfnamefont {S.}~\bibnamefont {Lin}}, \bibinfo {author} {\bibfnamefont
  {W.}~\bibnamefont {Lu}}, \bibinfo {author} {\bibfnamefont {X.}~\bibnamefont
  {Zhu}}, \bibinfo {author} {\bibfnamefont {Z.}~\bibnamefont {Yang}}, \bibinfo
  {author} {\bibfnamefont {W.}~\bibnamefont {Song}}, \bibinfo {author}
  {\bibfnamefont {J.}~\bibnamefont {Dai}}, \ and\ \bibinfo {author}
  {\bibfnamefont {Y.}~\bibnamefont {Sun}},\ }\href {\doibase
  http://dx.doi.org/10.1016/j.scriptamat.2011.05.035} {\bibfield  {journal}
  {\bibinfo  {journal} {Scripta Materialia}\ }\textbf {\bibinfo {volume}
  {65}},\ \bibinfo {pages} {452 } (\bibinfo {year} {2011})}\BibitemShut
  {NoStop}%
\bibitem [{\citenamefont {Szunyogh}\ \emph {et~al.}(2009)\citenamefont
  {Szunyogh}, \citenamefont {Lazarovits}, \citenamefont {Udvardi},
  \citenamefont {Jackson},\ and\ \citenamefont {Nowak}}]{szunyogh-2009-79}%
  \BibitemOpen
  \bibfield  {author} {\bibinfo {author} {\bibfnamefont {L.}~\bibnamefont
  {Szunyogh}}, \bibinfo {author} {\bibfnamefont {B.}~\bibnamefont
  {Lazarovits}}, \bibinfo {author} {\bibfnamefont {L.}~\bibnamefont {Udvardi}},
  \bibinfo {author} {\bibfnamefont {J.}~\bibnamefont {Jackson}}, \ and\
  \bibinfo {author} {\bibfnamefont {U.}~\bibnamefont {Nowak}},\ }\href
  {\doibase 10.1103/PhysRevB.79.020403} {\bibfield  {journal} {\bibinfo
  {journal} {Phys. Rev. B}\ }\textbf {\bibinfo {volume} {79}},\ \bibinfo
  {pages} {020403(R)} (\bibinfo {year} {2009})}\BibitemShut {NoStop}%
\bibitem [{\citenamefont {Kohn}\ \emph {et~al.}(2013)\citenamefont {Kohn},
  \citenamefont {Kovacs}, \citenamefont {Fan}, \citenamefont {McIntyre},
  \citenamefont {Ward},\ and\ \citenamefont {Goff}}]{Kohn:2013}%
  \BibitemOpen
  \bibfield  {author} {\bibinfo {author} {\bibfnamefont {A.}~\bibnamefont
  {Kohn}}, \bibinfo {author} {\bibfnamefont {A.}~\bibnamefont {Kovacs}},
  \bibinfo {author} {\bibfnamefont {R.}~\bibnamefont {Fan}}, \bibinfo {author}
  {\bibfnamefont {G.~J.}\ \bibnamefont {McIntyre}}, \bibinfo {author}
  {\bibfnamefont {R.~C.~C.}\ \bibnamefont {Ward}}, \ and\ \bibinfo {author}
  {\bibfnamefont {J.~P.}\ \bibnamefont {Goff}},\ }\href@noop {} {\bibfield
  {journal} {\bibinfo  {journal} {Sci. Rep.}\ } (\bibinfo {year}
  {2013})}\BibitemShut {NoStop}%
\bibitem [{\citenamefont {Rai{\v{c}}evi{\'{c}}}\ \emph
  {et~al.}(2011)\citenamefont {Rai{\v{c}}evi{\'{c}}}, \citenamefont
  {Popovi{\'{c}}}, \citenamefont {Panagopoulos}, \citenamefont {Benfatto},
  \citenamefont {Silva~Neto}, \citenamefont {Choi},\ and\ \citenamefont
  {Sasagawa}}]{Raicev:PhysRevLett.106.227206}%
  \BibitemOpen
  \bibfield  {author} {\bibinfo {author} {\bibfnamefont {I.}~\bibnamefont
  {Rai{\v{c}}evi{\'{c}}}}, \bibinfo {author} {\bibfnamefont {D.}~\bibnamefont
  {Popovi{\'{c}}}}, \bibinfo {author} {\bibfnamefont {C.}~\bibnamefont
  {Panagopoulos}}, \bibinfo {author} {\bibfnamefont {L.}~\bibnamefont
  {Benfatto}}, \bibinfo {author} {\bibfnamefont {M.~B.}\ \bibnamefont
  {Silva~Neto}}, \bibinfo {author} {\bibfnamefont {E.~S.}\ \bibnamefont
  {Choi}}, \ and\ \bibinfo {author} {\bibfnamefont {T.}~\bibnamefont
  {Sasagawa}},\ }\href {\doibase 10.1103/PhysRevLett.106.227206} {\bibfield
  {journal} {\bibinfo  {journal} {Phys. Rev. Lett.}\ }\textbf {\bibinfo
  {volume} {106}},\ \bibinfo {pages} {227206} (\bibinfo {year}
  {2011})}\BibitemShut {NoStop}%
\bibitem [{\citenamefont {Kim}\ \emph {et~al.}(2010)\citenamefont {Kim},
  \citenamefont {Chang}, \citenamefont {Korenblit}, \citenamefont {Islam},
  \citenamefont {Edwards}, \citenamefont {Freericks}, \citenamefont {Duan},\
  and\ \citenamefont {Monroe}}]{Kim:nature_2010}%
  \BibitemOpen
  \bibfield  {author} {\bibinfo {author} {\bibfnamefont {K.}~\bibnamefont
  {Kim}}, \bibinfo {author} {\bibfnamefont {M.-S.}\ \bibnamefont {Chang}},
  \bibinfo {author} {\bibfnamefont {S.}~\bibnamefont {Korenblit}}, \bibinfo
  {author} {\bibfnamefont {R.}~\bibnamefont {Islam}}, \bibinfo {author}
  {\bibfnamefont {E.~E.}\ \bibnamefont {Edwards}}, \bibinfo {author}
  {\bibfnamefont {G.-D.}\ \bibnamefont {Freericks}, \bibfnamefont
  {J.~K.and~Lin}}, \bibinfo {author} {\bibfnamefont {L.-M.}\ \bibnamefont
  {Duan}}, \ and\ \bibinfo {author} {\bibfnamefont {C.}~\bibnamefont
  {Monroe}},\ }\href {\doibase 10.1038/nature09071} {\bibfield  {journal}
  {\bibinfo  {journal} {Nature}\ }\textbf {\bibinfo {volume} {465}},\ \bibinfo
  {pages} {590} (\bibinfo {year} {2010})}\BibitemShut {NoStop}%
\bibitem [{\citenamefont {Tatara}\ and\ \citenamefont
  {Fukuyama}(1997)}]{Tatara_PhysRevLett.78.3773}%
  \BibitemOpen
  \bibfield  {author} {\bibinfo {author} {\bibfnamefont {G.}~\bibnamefont
  {Tatara}}\ and\ \bibinfo {author} {\bibfnamefont {H.}~\bibnamefont
  {Fukuyama}},\ }\href {\doibase 10.1103/PhysRevLett.78.3773} {\bibfield
  {journal} {\bibinfo  {journal} {Phys. Rev. Lett.}\ }\textbf {\bibinfo
  {volume} {78}},\ \bibinfo {pages} {3773} (\bibinfo {year}
  {1997})}\BibitemShut {NoStop}%
\end{thebibliography}
%

\end{document}